# Modeling Heterogeneous Mediation Effects in Survival Analysis via an Interpretable M-Learner Framework

**Xingyu Li[1], Qing Liu[2], Xun Jiang[2], Hong Amy Xia[2], Brian P. Hobbs[3], Peng Wei[1]***

1. Department of Biostatistics, The University of Texas MD Anderson Cancer Center, Houston, TX, 77030, USA.

2. Center for Design and Analysis, Amgen, Thousand Oaks, CA, 91320, USA.

3. Telperian Inc, TX, 78738, USA.

**email:* PWei2@mdanderson.org

SUMMARY: Mediation analysis is a useful tool to evaluate surrogate endpoints in clinical trials. We propose a novel method, the M-survival learner, for estimating heterogeneous indirect treatment effects in the presence of censored outcomes. The proposed approach enables the identification of interpretable patient subgroups characterized by distinct mediation pathways. To distinguish heterogeneous from homogeneous mediation effects, we introduce a new statistical criterion specifically designed for survival data. The method provides a principled framework for evaluating heterogeneity in surrogate biomarker performance across patient populations, offering evidence to support accelerated approval drug. By explicitly assessing subgroup-specific surrogate validity, the proposed approach addresses key regulatory concerns regarding the reliability of surrogate endpoints. We further establish theoretical properties of the method to justify its statistical guarantees. We apply the approach to data from a Phase III randomized clinical trial of HIV treatment, demonstrating its practical utility in real-world settings. Extensive simulation studies further evaluate and demonstrate its finite-sample performance.

## 1. Introduction

Evaluating the clinical benefit of a new therapeutic intervention, defined by improvements in survival, symptoms, or functional outcomes, often requires prolonged follow-up, particularly in chronic or life-threatening diseases. To address the inherent delays associated with measuring such endpoints, the U.S. Food and Drug Administration (FDA) established the Accelerated Approval pathway in 1992 (FDA, 1992). This regulatory mechanism permits approval of drugs for serious conditions with unmet medical needs on the basis of surrogate or intermediate clinical endpoints, thereby expediting patient access to potentially effective therapies.

A surrogate endpoint is a biomarker, such as a laboratory measurement, radiographic finding, or physical sign, that is intended to predict clinical benefit but does not itself directly measure it (FDA, 2014). An intermediate clinical endpoint similarly reflects a therapeutic effect that is reasonably likely to forecast ultimate clinical benefit, for example through its influence on irreversible morbidity or mortality. By substituting such endpoints for definitive outcomes, the accelerated approval framework can substantially shorten the drug development timeline. In oncology, for instance, regulatory approval may be granted based on tumor response rather than overall survival (OS), under the assumption that tumor shrinkage is reasonably likely to predict meaningful clinical benefit.

Despite widespread reliance on surrogate endpoints, existing evaluation frameworks implicitly treat surrogacy as a homogeneous, population-level property, an assumption that is rarely scrutinized and may be untenable in the presence of biological and clinical heterogeneity. As a result, a fundamental gap remains in our ability to characterize when and for whom a surrogate reliably predicts clinical benefit.

Surrogate endpoints introduces substantial challenges. Post-approval confirmatory trials are required to verify that treatment effects on the surrogate translate into true clinical benefit, for example, OS, yet an increasing number of such trials have failed to do so. To date, the FDA has withdrawn accelerated approval for 31 oncology drugs, 25 of them after 2020, as well as nine additional drugs in non-oncology indications. In response, between 2023 and 2025 the FDA issued multiple draft guidances that strengthened oversight of accelerated approvals based on surrogate biomarkers, including requirements that confirmatory trials be underway at the time of approval (FDA, 2023, 2024, 2025). These developments underscore growing concern regarding the validity and reliability of surrogate endpoints, particularly when treatment effects and surrogate–outcome relationships are heterogeneous.

The history of gefitinib illustrates the potential heterogeneity in surrogate performance across patient subgroups. In 2003, gefitinib received accelerated approval for patients with advanced non–small cell lung cancer based on an objective response rate (ORR) of approximately 10.6% in the overall population, where tumor response was considered a surrogate endpoint reasonably likely to predict clinical benefit (Cohen et al., 2003). However, subsequent confirmatory trials failed to demonstrate an OS benefit in the overall population, suggesting that tumor response did not reliably translate into improved survival at the population level. In contrast, when evaluated in patients with EGFR-mutant tumors, gefitinib achieved substantially higher response rates and clear survival benefits (Kazandjian et al., 2016). This example demonstrates that the relationship between surrogate endpoints and clinical outcomes may vary across patient subgroups, and that population-level analyses may obscure subgroup-specific mechanisms.

In practice, this assumption may fail when surrogate performance varies across individuals or subgroups, leading to inconsistencies between early-phase surrogate-based evidence and definitive clinical outcomes. Although these examples exhibit substantial treatment effect heterogeneity, our focus is on heterogeneity in surrogacy. Treatment heterogeneity alone does not invalidate a surrogate if improvements in the surrogate consistently translate into clinical benefit across populations. The concern arises when the relationship between surrogate improvement and survival benefit varies by subgroup or trial context. In such settings, surrogate validity itself becomes heterogeneous rather than a purely population-level property.

Together, these examples highlight a fundamental limitation of commonly used surrogate endpoints such as ORR: the surrogacy relationship between the surrogate and the clinical endpoint may differ across patient subgroups. These discrepancies motivate a shift from viewing surrogate validity as a global property toward explicitly modeling heterogeneity in treatment effects associated with the surrogate. Rather than treating surrogacy as a population-level characteristic, its scientific value lies in identifying the populations for which changes in a surrogate endpoint are predictive of survival benefit.

Beyond their role in regulatory approval, surrogate endpoints have broad implications for drug development and precision medicine. When appropriately validated, they can improve early-phase decision-making by providing timely signals of efficacy, enabling more informed "go/no-go" decisions and reducing the financial and temporal risks of late-stage trials.

Despite these potential advantages, both regulatory agencies and methodological researchers have increasingly emphasized the need for more rigorous statistical frameworks to assess surrogate validity, particularly in the presence of treatment effect heterogeneity. Existing surrogate validation approaches, including classical criteria-based methods, meta-analytic frameworks, and principal stratification–based approaches, primarily rely on population-level summaries or latent strata. As a result, they are not designed to uncover heterogeneity in surrogate-mediated treatment effects at the individual or subgroup level (Li et al., 2011; Zhou et al., 2021).

To address this limitation, several recent studies have proposed covariate-adjusted causal mediation frameworks. For example, Parast et al. (2023) introduced a covariate-adjusted approach that evaluates surrogate validity using the proportion of treatment effect explained by the surrogate conditional on baseline characteristics. This framework was later extended by Parast et al. (2024) to accommodate survival outcomes. However, these approaches are limited in their ability to detect complex forms of heterogeneity in surrogate effects. Building on these methods, Knowlton et al. (2025) proposed a grid-based procedure that probes the covariate space to identify regions where surrogate strength exceeds a clinically meaningful threshold while controlling for multiplicity. Although this approach allows for more flexible detection of heterogeneous regions, it becomes computationally challenging as the number of covariates increases and may struggle to identify regions with complex boundaries.

These limitations motivate the development of the M-survival learner. The primary objective of this approach is to characterize heterogeneity in surrogate-mediated treatment effects and to evaluate surrogate validity across clinically relevant subgroups. By explicitly modeling individual-level variation in indirect treatment effects, the proposed framework aims to identify patient populations in which the surrogate reliably mediates the treatment effect on the clinical endpoint. This subgroup-level perspective not only provides a more nuanced assessment of surrogate validity but also generates actionable evidence that can inform regulatory decision-making within the FDA's accelerated approval pathway.

Specifically, our contributions are: we formalize surrogate validity as a heterogeneous, individual-level property rather than a global characteristic. We develop a learner-based (flexible machine learning method, such as XGboost and neural network) framework to estimate indirect treatment effects on survival outcome. We introduce a clustering and profiling strategy to identify clinically interpretable subgroups with distinct surrogate-mediated effects.

The remainder of this article is organized as follows. In Section 2, we introduce the surrogate biomarker and mediation framework and describe their conceptual relationship. Section 3 presents a heterogeneous mediation model that allows indirect treatment effects to vary across individuals. In Section 4, we develop a clustering strategy to identify subgroups with distinct surrogate effects, and introduces an interpretable profiling approach to characterize these subgroups. Section 5 evaluates finite-sample performance under a range of scenarios, and Section 6 illustrates the proposed methods using data from a phase III randomized clinical trial on HIV. Finally, Section 7 summarizes our findings and discusses their implications and potential extensions.

## 2. Surrogate Biomarker and Mediation Model

This section provides the conceptual foundation for the proposed methodology by reviewing surrogate biomarkers and mediation analysis, and clarifying their connections from a causal inference perspective. We first introduce surrogate endpoints and discuss ideal and non-ideal surrogacy through causal diagrams. We then review existing approaches to surrogate validation and highlight their limitations in the presence of heterogeneity, thereby motivating the need for heterogeneous mediation models.

### 2.1 *Surrogate Biomarkers*

Surrogate endpoints are post-treatment variables intended to capture the effect of an intervention on a clinically meaningful outcome, particularly in settings where definitive endpoints such as OS are rare, delayed, or costly to observe. By leveraging treatment effects on surrogate endpoints, investigators aim to infer treatment effects on the true clinical outcome, thereby accelerating evaluation and decision-making in clinical trials (Buyse et al., 2010). Surrogate endpoints have played a prominent role in regulatory science; for example, pathological complete response has been used in breast cancer, and durable ORR has frequently served as a surrogate endpoint for regulatory approval in solid tumors (FDA, 2025).

From a causal perspective, surrogate endpoints are most naturally viewed as mediators lying on the pathway between treatment and the clinical outcome. Figure 1 illustrates two stylized causal structures. In panel (a), the surrogate endpoint $M$ represents an ideal surrogate, in which the entire causal effect of treatment on the clinical outcome $T$ is mediated through $M$, and there is no residual direct effect of treatment on $T$ once $M$ is controlled for. A canonical example arises in chronic myeloid leukemia, where levels of BCR-ABL transcripts serve as a molecular marker of disease burden (FDA, 2021). Tyrosine kinase inhibitors, such as imatinib, are designed to directly inhibit the BCR–ABL fusion protein, and reductions in BCR–ABL transcript levels closely reflect the biological mechanism of treatment action. As a result, molecular response has been widely accepted as a surrogate for long-term clinical outcomes, including progression-free and overall survival, particularly in early-phase trials.

[Figure 1 about here.]

In contrast, most surrogate endpoints encountered in practice are non-ideal. Panel (b) of Figure 1 depicts a setting in which the surrogate captures only part of the treatment effect on the clinical outcome, with a remaining direct effect of treatment acting independently of the surrogate. Tumor response provides a paradigmatic example of such an imperfect surrogate. Although reductions in tumor burden indicate biological activity of treatment, they often fail to fully account for downstream effects on survival, toxicity, or disease progression, resulting in substantial residual treatment effects that are not mediated by the surrogate.

### 2.2 *Existing Approaches and the Challenge of Heterogeneity*

A large body of methodological literature has developed to assess the validity of surrogate biomarkers. Most existing approaches, however, focus on average effects at the population or latent stratum level, implicitly assuming homogeneous surrogate validity across patients (Li et al., 2011). Prominent examples include principal stratification based methods, which define surrogate-related causal effects within latent subgroups characterized by potential outcomes; meta-analytic approaches, which aggregate surrogate outcome associations across multiple trials to evaluate population-level surrogacy; and mediation analysis methods, such as the proportion-explained framework originally proposed by Freedman (Freedman and Schatzkin, 1992). While these frameworks provide valuable insights into overall or latent surrogate effects, they are not designed to characterize how surrogate-mediated treatment effects vary across individuals or clinically defined subpopulations.

These findings highlight a fundamental limitation of population-level surrogate validation. Even when a biomarker satisfies commonly used criteria for surrogate validity on average, its reliability as a surrogate may vary substantially across individuals or subgroups, reflecting heterogeneity in the underlying causal pathways linking treatment, biomarker, and clinical outcome. Reliance on population-level validation alone may therefore lead to erroneous decisions in early trial termination, accelerated regulatory approval, or individualized clinical decision-making. These concerns have contributed to increasing caution among regulatory agencies, including the FDA, regarding the use of surrogate endpoints in settings characterized by biological and clinical heterogeneity.

Together, these considerations motivate the development of methodological frameworks that explicitly account for heterogeneity in surrogate validity. In particular, there is a need for approaches that characterize individual- and subgroup-level variation in surrogate-mediated treatment effects, rather than relying solely on population-average measures. Our framework is based on heterogeneous mediation analysis. Xue et al. (2022); Li et al. (2025) investigate heterogeneity in mediation effects. However, these approaches have not been developed for survival outcomes.

Section 3 introduces a heterogeneous mediation framework designed to address this need by estimating individual-level surrogate-mediated treatment effects. Section 4 presents a clustering-based approach for identifying subgroups with distinct surrogate validity profiles.

Taken together, these components constitute the proposed methodology. Figure 1 (d) illustrates the overall analytical pipeline of the proposed approach.

## 3. Heterogeneous Mediation Treatment Effects

In this section, we characterize heterogeneity in treatment effects transmitted through a surrogate biomarker by modeling heterogeneous mediation effects within a causal mediation framework.

We consider a randomized controlled trial (RCT) with $n$ independent and identically distributed individuals. For notational simplicity, we suppress the individual index throughout and work with generic random variables. Let $W \in \{0, 1\}$ denote the treatment indicator, and let $\boldsymbol{X} \in \mathbb{R}^p$ denote a vector of baseline covariates taking values in a covariate space $\mathcal{X} \subseteq \mathbb{R}^p$. Let $M \in \mathbb{R}$ denote a post-treatment mediator with support $\mathcal{M} \subseteq \mathbb{R}$.

The primary outcome of interest is a time-to-event variable $T \in \mathbb{R}_+$, and let $C \in \mathbb{R}_+$ denote the censoring time. Due to right censoring, the observed time-to-event data are represented by $(U, \delta)$, where $U = \min(T, C)$ and $\delta = \mathbb{I}(T \leqslant C)$. Throughout this paper, we assume that the mediator is measured prior to the occurrence of the event or censoring.

Within the potential outcomes framework, let $M(w)$ denote the mediator value that would be observed under treatment level $w$. Let $T(w, m)$ denote the potential event time that would have been observed if, possibly contrary to fact, treatment were set to $w$ and the mediator were set to level $m \in \mathcal{M}$. For example, $T(0, M(1))$ represents the event time that would have been observed had the individual not received treatment while the mediator were set to the level it would have attained under treatment.

We link the potential outcomes to the observed data through the consistency assumption, which states that we observe the mediator $M = M(W)$, $T = T(W, M(W))$. Moreover, because the potential mediator values and potential outcomes are defined solely as functions of an individual's own treatment assignment, and not of the treatment assignments received by other individuals, this formulation also implicitly rules out interference between units. This condition is commonly referred to as the Stable Unit Treatment Value Assumption (SUTVA) (Rubin et al., 1990).

### 3.1 *Identification*

Following Imai et al. (2010), we make the following sequential ignorability assumption throught, allowing for the identification of the natural direct and indirect treatment effects.

ASSUMPTION 1 (Randomization): In a randomized controlled trial, the treatment indicator $W$ is independent of baseline covariates $\boldsymbol{X}$, that is,

$$W \perp\!\!\!\perp \boldsymbol{X}, \qquad \boldsymbol{X} \in \mathcal{X}.$$

ASSUMPTION 2 (Ignorability of treatment assignment): Conditional on baseline covariates $\boldsymbol{X}$, the treatment assignment is ignorable with respect to the potential mediator and potential outcomes, that is,

$$\{T(w^*, M(w)), M(w)\} \perp\!\!\!\perp W \mid \boldsymbol{X}, \quad w, w^* \in \{0, 1\}.$$

ASSUMPTION 3 (Ignorability of the mediator): Conditional on the observed treatment assignment and baseline covariates, the mediator is ignorable with respect to the potential outcomes, that is,

$$T(w^*, M(w)) \perp\!\!\!\perp M \mid W = w, \boldsymbol{X}, \quad w, w^* \in \{0, 1\}.$$

ASSUMPTION 4 (Positivity): We assume positivity of the mediator, such that

$$f_M\{M(w) = m \mid W = w, \boldsymbol{X}\} > 0, \quad \text{for all } m \in \mathcal{M},\ w \in \{0, 1\},\ \boldsymbol{X} \in \mathcal{X}.$$

The first assumption formalizes the randomization mechanism of the treatment assignment in an RCT, ensuring that the treatment indicator is independent of baseline covariates. This assumption reflects the study design and is typically guaranteed by proper randomization.

The second assumption states that, conditional on baseline covariates, the treatment assignment is ignorable with respect to the potential mediator and potential outcomes. This assumption is automatically satisfied in randomized experiments and allows the treatment effects on both the mediator and the primary outcome to be identified.

The third assumption requires ignorability of the mediator given the observed treatment and baseline covariates, namely that the mediator is independent of the potential outcomes after conditioning on treatment and covariates. Unlike the ignorability of treatment assignment, this assumption is generally not guaranteed even in randomized experiments, as the mediator is not randomly assigned. Consequently, this assumption is not directly testable from the observed data and must be justified based on substantive knowledge and careful study design.

The fourth assumption is a positivity condition for the treatment and the mediator, requiring that each treatment level and a sufficient range of mediator values occur with positive probability within each covariate stratum. This assumption ensures that the causal mediation effects are identifiable from the observed data and can be consistently estimated.

Many existing mediation analysis methods for time-to-event outcomes under the Cox proportional hazards model define the mediation effect on the log-hazard scale (VanderWeele, 2011), we define the conditional direct and indirect treatment effects on the log-hazard scale under a Cox proportional hazards framework. The outcome model is specified as

$$\lambda(t|\boldsymbol{X}, W, M) = \lambda_0(t) exp(g(\boldsymbol{X}, W, M)), \tag{1}$$

where

$$g(\boldsymbol{X}, W, M) = \kappa_1(\boldsymbol{X}) + \kappa_2(\boldsymbol{X}) \cdot W + \kappa_3(\boldsymbol{X}) \cdot M. \tag{2}$$

Here, $\lambda_0(t)$ denotes the baseline hazard function. The function $\kappa_1(\boldsymbol{X})$ captures the effect of baseline covariates, $\kappa_2(\boldsymbol{X})$ represents the treatment effect on the outcome not operating through the mediator, and $\kappa_3(\boldsymbol{X})$ is a covariate-dependent coefficient characterizing effect modification of the mediator on the hazard. When $\kappa_1(\boldsymbol{X})$ and $\kappa_2(\boldsymbol{X})$ are linear in $\boldsymbol{X}$ and $\kappa_3(\boldsymbol{X})$ is constant, the model corresponds to a linear assumption. More generally, these functions are allowed to be general functions with finite expectations, accommodating flexible and potentially nonlinear relationships, we call it complex assumption; more disscussion can be found in Appendix A.

The mediator is modeled as

$$M = \kappa_4(\boldsymbol{X}) + \kappa_5(\boldsymbol{X}) \cdot W + \epsilon, \tag{3}$$

where $\kappa_4(\boldsymbol{X})$ represents the baseline covariate effects on the mediator, $\kappa_5(\boldsymbol{X})$ denotes the treatment effect on the mediator, and $\epsilon$ is a mean-zero term. Both $\kappa_4(\boldsymbol{X})$ and $\kappa_5(\boldsymbol{X})$ can be nonlinear functions. Under these notations, the total treatment effects (TTE) and natural direct and indirect treatment effects (DTE, ITE) on log hazard scale conditional on covariates are given by,

$$\begin{aligned}
TTE(t|\boldsymbol{X}=\boldsymbol{x}) &:= \log\{\lambda_{T_{w,M_w}}(t|\boldsymbol{X}=\boldsymbol{x})\} - \log\{\lambda_{T_{w^*,M_{w^*}}}(t|\boldsymbol{X}=\boldsymbol{x})\},\\
ITE(t|\boldsymbol{X}=\boldsymbol{x}) &:= \log\{\lambda_{T_{w,M_w}}(t|\boldsymbol{X}=\boldsymbol{x})\} - \log\{\lambda_{T_{w,M_{w^*}}}(t|\boldsymbol{X}=\boldsymbol{x})\},\\
DTE(t|\boldsymbol{X}=\boldsymbol{x}) &:= \log\{\lambda_{T_{w,M_{w^*}}}(t|\boldsymbol{X}=\boldsymbol{x})\} - \log\{\lambda_{T_{w^*,M_{w^*}}}(t|\boldsymbol{X}=\boldsymbol{x})\}.
\end{aligned} \tag{4}$$

For simplicity of notation, $T_{w,M_w}$ is denoted as $T(w, M(w))$, counterfactual log hazard conditonal on covariates $\boldsymbol{X}=\boldsymbol{x}$ is denoted as $\log\{\lambda_{T_{w,M_w}}(t|\boldsymbol{X}=\boldsymbol{x})\}$.

Since we focus on the natural indirect treatment effects conditional on $W = 1$, we fix $w = 1$ in the outcome model and set $w^* = 0$ in the mediator distribution. Substituting $w = 1$ and $w^* = 0$ into the expression (4) therefore yields the indirect treatment effect considered throughout this paper.

THEOREM 1 (Identification): *Under Assumption 1-4 with rare event assumption, the indirect treatment effect conditional on covariates could be identified as*

$$\begin{aligned}
ITE(t|\boldsymbol{X}=\boldsymbol{x}) = \kappa_3(\boldsymbol{x})\kappa_5(\boldsymbol{x}) = g(\boldsymbol{x}|W=1, M=\mathrm{E}[M|w=1, \boldsymbol{X}=\boldsymbol{x}]) - \\
g(\boldsymbol{x}|W=1, M=\mathrm{E}[M|w=0, \boldsymbol{X}=\boldsymbol{x}]).
\end{aligned} \tag{5}$$

REMARK 1: Under Assumptions 1–4 and the rare event assumption, $ITE(t|\boldsymbol{X})$, which we refer to as the natural indirect treatment effect conditional on covariates (NIECC), admits a particularly simple form. Specifically, although the NIECC is defined at time $t$, its final expression does not depend on $t$. The NIECC is given by the product $\kappa_3(\boldsymbol{x})\kappa_5(\boldsymbol{x})$, which captures the effect of the mediator on the outcome and the effect of the treatment on the mediator, respectively. Equivalently, the same quantity can be expressed as the difference in the outcome regression function $g(\cdot)$ evaluated at the expected mediator values under $W = 1$ and $W = 0$, conditional on $\boldsymbol{X} = \boldsymbol{x}$.

REMARK 2: To illustrate heterogeneous mediation effects, we provide an example in Figure 1. In this setting, patients in Group 1 have only ITE, patients in Group 2 have both ITE and DTE , whereas patients in Groups 3 exhibit only DTE.

The NIECC admits a particularly simple representation. Proof of Theorem 1 is provided in Appendix D and additional discussion and details of the estimation procedure are provided in Appendix A.

## 4. Clustering and Identify Patient Profiles

### 4.1 *Clustering based on similarity*

We quantify the treatment effect transmitted through the surrogate using a heterogeneous mediation analysis framework that allows the NIECC to vary with baseline covariates. In many clinical settings, the validity and strength of a surrogate biomarker are not homogeneous across the study population but instead depend critically on patient characteristics. Consequently, a surrogate biomarker may be informative only within specific regions of the covariate space, exhibiting substantial heterogeneity across individuals. From a practical perspective, it is therefore important to identify interpretable covariate-defined regions or subgroups in which the surrogate consistently captures the treatment effect, as well as regions in which its performance differs or deteriorates. This task can be viewed as the identification of surrogate-relevant subgroups, characterized by relatively homogeneous indirect treatment effects within subgroups and substantial heterogeneity across subgroups.

Motivated by this objective, we propose a novel similarity-based clustering approach built upon a modified t-SNE representation combined with K-means clustering. t-SNE is a dimensionality reduction algorithm that preserves similarity among observations by modeling pairwise distances and projecting the data into a lower-dimensional space. In our framework, the pairwise distance between observations is defined based on the NIECC. Detailed descriptions of the method and the corresponding theoretical analyses are provided in Appendix B, E, and F.

### 4.2 *Identify Patient Profiles*

Interpretability is crucial for translating statistical patterns into clinical practice. Machine learning models, when applied in clinical research, should ideally yield interpretable results that are both understandable and logically sound for practicing clinicians. Although accurate outcome prediction is necessary, it is not sufficient for developing effective models to guide treatment selection. Owing to the properties of t-SNE, similar observations tend to concentrate near the centers of clusters in the embedded space. Consequently, we apply K-means clustering to identify concrete subgroups. Decision trees offer a highly interpretable framework for clinical decision making that is easy to disseminate through structured diagrams depicting a step-by-step sequence of variable assessments. Assessing a tree’s alignment with clinical reasoning is straightforward for clinicians, as it does not require an understanding of the potentially complex modeling processes used to generate it.

Both t-SNE and K-means involve random initialization and tuning parameters. In particular, K-means requires the number of clusters to be specified within a predefined range and is sensitive to this choice, which can lead to variability in the resulting patient profiles across repeated runs. To address this issue, we propose a custom metric to identify the most representative patient profile among the solutions obtained under different clustering specifications. We make the following rules to select the final clustering result: For each pre-defined $k$ clusters, we note them as $C_1, \cdots, C_k$.

Step 1: We use a decision tree to fit $k$ clusters from the K-means clustering, denoting each terminal node of the decision tree leaf.

Step 2: Two Cox models are fitted,

$$\lambda(t) = \lambda_0(t)\exp\{\beta_1 W\}, \tag{6}$$

$$\lambda(t) = \lambda_0(t)\exp\{\beta_2 W + \beta_3 \text{leaf} + \beta_4 \text{leaf} * W\}. \tag{7}$$

Let $l_1$ and $l_2$ denote the maximized partial log-likelihoods of models (6) and (7), respectively. The likelihood ratio test statistic,

$$-2(l_2 - l_1) \rightarrow \chi^2(df), \tag{8}$$

where $df$ = number of leaves −1. $p^{Y}_{\text{leaf}}$ is the p value of model (8).

Step 3: Two linear models are fitted,

$$M = \beta_5 W, \tag{9}$$

$$M = \beta_6 W + \beta_7 \text{leaf} + \beta_8 \text{leaf} * W. \tag{10}$$

Let $l_3$ and $l_4$ denote the maximized likelihoods of models (9) and (10), respectively. The likelihood ratio test statistic,

$$-2(l_2 - l_1) \rightarrow \chi^2(df), \tag{11}$$

where $df$ = number of leaves −1. $p^{M}_{\text{leaf}}$ is the p value of model (11)

Step 4: The selection metric is:

$$metric = I(p^{Y}_{\text{leaf}} < p^{Y^*}) \times I(p^{M}_{\text{leaf}} < p^{M^*}) \times p^{M}_{\text{leaf}}, \tag{12}$$

where $p^{Y^*}$ and $p^{M^*}$ serve as the thresholds for determining the presence of heterogeneity in the data and are selected through a calibration procedure. We select the profile with the smallest metric as the final output. In simulation studies, $p_{\text{leaf}}$ is calibrated under the non-heterogeneous scenario to control the Type I error rate. In practical applications, the corresponding threshold can be determined via a permutation-based procedure. More details are provided in Appendix C.

## 5. Simulation Study

This Section presents a comprehensive evaluation of the proposed method via a series of simulation studies. We also describe the calibration procedure incorporated in the proposed approach. The simulation study consisted of two parts. The first part is linear assumption. The second part is complex assumption. In the following, we describe seperately.

### 5.1 *Linear assumption*

The OS outcomes and mediators were generated according to the following models:

$$\begin{aligned} \lambda(t|M, W, \boldsymbol{X}) &= \lambda_0(t) \times \exp\{\sum_{j=1}^{10} \beta_j X^{(j)} + \beta_w W + \beta_m M\}, \\ M(W, X) &= \kappa_1(\boldsymbol{X}) + \kappa_2(\boldsymbol{X}) W + \epsilon, \end{aligned} \tag{13}$$

where $\lambda(t|M, W, \boldsymbol{X})$ denoted the hazard function of the survival time conditional on the mediator $M$, treatment assignment $W$, and baseline $\boldsymbol{X}$. The baseline hazard function $\lambda_0(t) = vt^{v-1}/\lambda^v$ followed a Weibull distribution with shape parameter $v$ and scale parameter $\lambda$, given by $\lambda_0(t) = vt^{v-1}/\lambda^v$. The true parameter values were set to $v = 2$ and $\lambda = 1/300$.

The covariate vector $\boldsymbol{X} = (X^{(1)}, \cdots, X^{(10)})$ consisted of 10 baseline covariates, where each $X^{(j)}$ was independently generated from a standard normal distribution. The coefficients $\beta_1, \cdots, \beta_{10}$ corresponded to the effects of the baseline covariates, $\beta_m$ represented the coefficient of mediator, and $\beta_w$ represented the treatment effect associated with the binary treatment indicator $W$. Subjects were randomly assigned to the treatment group ($W = 1$) or the control group ($W = 0$) in a $1 : 1$ ratio, and the sample size was 1000. In the mediator model, $\kappa_1(\boldsymbol{X})$ represented the non-treatment-related component of the mediator, while $\kappa_2(\boldsymbol{X})$ captured treatment–covariate interactions and thus induced treatment effect heterogeneity. The error term $\epsilon$ was generated independently to represent unexplained variability in the mediator. We considered three simulation scenarios: Heterogeneous, Global, and Null. In the Heterogeneous setting, the mediator mediated the treatment effect only in certain regions of the covariate space; in the Global setting, it mediated the treatment effect for all individuals; and in the Null setting, the surrogate did not mediated the treatment effect in any subgroup. Additional experimental details and full specifications of each scenario were provided in Appendix G.

5.2 *Complex assumption*

The OS outcomes, mediators were generated according to the following models:

$$\begin{aligned} \lambda(t|M, W, \boldsymbol{X}) &= \lambda_0(t) \times \exp\{\kappa_1(\boldsymbol{X}) + \kappa_2(\boldsymbol{X}) \cdot W + \kappa_3(\boldsymbol{X}) \cdot M\}, \\ M &= \kappa_4(\boldsymbol{X}) + \kappa_5(\boldsymbol{X}) \cdot W + \epsilon, \end{aligned} \tag{14}$$

where $\kappa_1(\boldsymbol{X})$ denoted the effects of baseline covariates, $\kappa_2(\boldsymbol{X})$ denoted the treatment effects not via the mediator, and $\kappa_3(\boldsymbol{X})$ was the varying coefficient on the mediator. In the mediator model, $\kappa_4(\boldsymbol{X})$ represented the non-treatment-related component of the mediator, while $\kappa_5(\boldsymbol{X})$ captured treatment–covariate interactions. Other settings were similar to Subsection 5.1.

We considered six simulation scenarios: All1, Part1, All2, Part2, Global, and Null. In the first four scenarios (All1, Part1, All2, and Part2), mediator validity held only within certain regions of the covariate space. The labels All and Part indicated whether the treatment effect was fully or partially mediated through the mediators, respectively. In All1 and Part1, the treatment effect was constant and did not vary with covariates, whereas in All2 and Part2, the treatment effect varied as a function of covariates. In the Global scenario, the mediator reflected the treatment effect for all individuals, while in the Null scenario, the mediator did not reflect the treatment effect in any subgroup. Detailed specifications of each scenario were provided in the Appendix G.

### 5.3 *Simulation Results*

All results under the linear model assumption were provided in Appendix G. Under the complex assumption, Figure 2 illustrated the selected threshold used for subgroup identification. The figure showed that the proposed method successfully identified regions with heterogeneous mediation effects, effectively distinguishing subgroups with distinct mediation patterns. Figure 3 showed the empirical cumulative distribution function of $p_{leaf}$, the results showed the ability of the proposed metric to distinguish the heterogeneous scenarios and homogeneous scenarios. The estimation accuracy of NIECC was evaluated in Appendix G, Web Table 1. More discussion and additional details regarding covariate selection and the exclusion of cases following calibration were presented in Appendix G. These results suggested that the proposed method was able to identify regions exhibiting heterogeneity in mediation treatment effects.

[Figure 2 about here.]

[Figure 3 about here.]

## 6. Real Data Application: The ACTG175 Trial

ACTG 175 (Hammer et al., 1996) is one of the earliest randomized clinical trials to demonstrate the superiority of combination therapy with zidovudine and didanosine over monotherapy with either zidovudine or didanosine among patients infected with human immunodeficiency virus (HIV). In this trial, each patient was randomly assigned to a specific treatment regimen. A key scientific question is whether a short-term endpoint measured several weeks after randomization can serve as a valid surrogate for a long-term clinical outcome, and whether the validity of such a surrogate may vary across different patient subpopulations.

In this study, we investigate these questions using the dataset “ACTG175” available in the R package “BART”, focusing on patients treated with zidovudine alone or the combination of zidovudine and didanosine.

The analysis adjusted for the following baseline covariates: age, body weight (wtkg), hemophilia status (hemo), homosexual activity (homo), Karnofsky performance score (karnof), prior antiretroviral therapy exposure measured as days of pre-ACTG 175 treatment (preanti), race, gender, symptomatic status, and baseline CD4 cell count (CD40). CD4 cell count measured at 20 weeks after randomization (±5 weeks) was selected as the surrogate biomarker, representing an early immunologic response relative to the median follow-up duration of 143 weeks. This surrogate precedes the clinical endpoint and captures short-term treatment effects. The primary endpoint was defined as a greater than 50% decline in CD4 cell count from baseline, progression to acquired immunodeficiency syndrome (AIDS), or death. For simplicity, this composite endpoint is hereafter referred to as OS.

Our analysis consisted of two parts. We first conducted the analysis under linear model assumptions and then extended the analysis to a more complex modeling framework. We presented only the results obtained under the complex assumptions, while the results and discussion based on the linear analysis were provided in Appendix H.

Subgroup 1 consisted of patients with low baseline CD4 levels and exhibited a substantially stronger mean NIECC compared with the other subgroups (Figure 4). Patients in this subgroup have poorer baseline immune status, and the treatment effect could be clearly mediated by the increase in CD4 counts at 20 weeks. This pattern suggested that CD4 served as an informative surrogate biomarker for treatment efficacy among patients with low baseline CD4 levels.

In contrast, Subgroup 2 and 3 both consisted of patients with relatively high baseline CD4 levels. Because these patients already had comparatively preserved immune function at baseline, the treatment effect mediated through CD4 improvement was considerably weaker. Nevertheless, meaningful differences remain between the two subgroups. Subgroup 3 included patients who had received antiretroviral treatment prior to enrollment (preanti $\geqslant$ 8 days), whereas Subgroup 2 consisted primarily of treatment-naïve patients. For patients in Subgroup 3, the treatment effect mediated through CD4 appeared to have a delayed impact. As shown in the counterfactual survival curves in Figure 5 (c), the three survival curves began to separate at later time points, indicating that the treatment effect mediated by CD4 improvement became more apparent over longer follow-up periods.

In contrast, for Subgroup 2, the curves for $S(1, M(1))$ and $S(1, M(0))$ almost completely overlap (Figure 5 (b)), suggesting that CD4 did not meaningfully capture the treatment effect in this subgroup. This observation is consistent with the distribution of individual NIECC shown in Figure 5 (a). Compared with Subgroup 2, Subgroup 3 exhibited greater heterogeneity in individual NIECC values, whereas Subgroup 2 showed a relatively concentrated distribution around zero, indicating a more stable but negligible mediation effect.

Overall, Figures 4 and 5 together revealed clinically interpretable heterogeneity in surrogate mediation effects and supported the validity of the identified subgroups.

[Figure 4 about here.]

[Figure 5 about here.]

## 7. Discussion

This work makes several methodological contributions to the evaluation of surrogate endpoints for survival outcomes. First, we formulate surrogate validity as a heterogeneous, individual-level property rather than a homogeneous population-level characteristic, allowing the degree to which a surrogate captures the treatment effect to vary across patients. Second, we develop a learner-based framework to estimate indirect treatment effects on survival out-comes, which accommodates flexible, potentially nonlinear relationships among treatment, surrogate, covariates, and time-to-event outcomes. Third, we introduce a clustering and profiling strategy to summarize this heterogeneity and to identify clinically interpretable subgroups characterized by distinct surrogate-mediated effects. Taken together, these innovations provide a flexible, data-adaptive framework for investigating heterogeneity in surrogate validity and for understanding when surrogate endpoints are more or less informative of treatment effects on survival.

Understanding heterogeneity in surrogate performance is particularly important in the context of clinical development, where surrogate endpoints are frequently used to support decision-making across the drug development continuum. The informativeness of a surrogate depends on multiple interacting factors, including the mechanism of action of the intervention, patient characteristics, disease etiology, and the definition of the downstream clinical endpoint. Consequently, the behavior of surrogate endpoints is often more complex than that of the clinical outcomes they are intended to predict. Population-average assessments of surrogate validity may therefore mask meaningful variation across patient subgroups.

These issues are especially salient in Phase II screening trials, which aim to determine whether a treatment demonstrates sufficient biological activity or preliminary efficacy to justify evaluation in a confirmatory Phase III trial. When direct assessment of the primary clinical outcome is impractical due to long follow-up times or limited sample sizes, surrogate endpoints such as tumor response, intermediate time-to-event outcomes, or biomarkers are often used to facilitate early go/no-go decisions. For example, in oncology trials, single-arm Phase II studies have historically relied on ORR, defined using RECIST criteria, as a primary endpoint for evaluating cytotoxic therapies (Zabor et al., 2020).

However, the therapeutic landscape has shifted markedly toward targeted and immunomodulatory agents with substantially more complex mechanisms of action. For such therapies, the relationship between early surrogate signals and long-term clinical benefit is often indirect, context-dependent, and heterogeneous across patient subgroups. As a result, surrogate endpoints that appear promising at the population level may fail to reliably predict downstream outcomes in confirmatory trials. These challenges have been reflected in recent regulatory actions, including the withdrawal of several drugs previously granted accelerated approval and the adoption of more stringent requirements for confirmatory evidence. A central driver of this increased scrutiny is the limited ability of existing evaluation frameworks to characterize how surrogate performance varies across patient populations and treatment contexts.

Most existing surrogate-based methods implicitly treat surrogate validity as a homogeneous, population-level property and do not explicitly model heterogeneity in the relationships among surrogate biomarkers, predictive biomarkers, and clinical outcomes. Consequently, these approaches rely on simplified assumptions that may be inadequate in modern therapeutic settings. Compared with Parast et al. (2024), our proposed method can directly identify heterogeneous regions of surrogate validity while simultaneously considering all covariates. In addition, our framework is grounded in a causal mediation analysis framework, which provides clearer interpretability for the surrogate-mediated treatment effects. Furthermore, our approach can detect more complex heterogeneous regions in the covariate space without requiring pre-specified grids to search for subgroup regions, allowing for a more flexible and data-adaptive identification of surrogate-valid subpopulations.

Importantly, the proposed approach is designed to distinguish genuine heterogeneity from spurious subgroup structure, thereby mitigating the risk of over-partitioning and false discoveries. By representing treatment effects along continuous dimensions of patient characteristics rather than relying on ad hoc subgroup definitions, the method yields a more interpretable and clinically meaningful characterization of heterogeneity. This representation has direct implications for precision treatment strategies and for the design of adaptive or subgroup-enriched clinical trials.

Application of the proposed method to the ACTG 175 HIV trial further illustrates its practical utility. The analysis revealed substantial heterogeneity primarily along baseline CD4 cell count and prior ART exposure. These findings are biologically plausible and consistent with established clinical knowledge: baseline CD4 count reflects immune reserve and capacity for immune reconstitution, while prior ART exposure captures treatment history, disease chronicity, and potential resistance, all of which are known to influence immunologic response to therapy (Dybul et al., 2002; Kaufmann et al., 2003; Clavel and Hance, 2004). The recovery of these well-recognized effect modifiers supports the validity of the proposed method and suggests that it captures clinically meaningful structure rather than noise.

From a practical perspective, these results indicate that the proposed framework can inform more precise treatment allocation and provide actionable insights for future trial design, including stratified or enrichment-based strategies. More broadly, systematic characterization of surrogate heterogeneity may help reconcile discrepancies between Phase II and Phase III findings, reduce uncertainty in early development decisions, and strengthen the evidentiary foundation underlying accelerated approval pathways.

Several limitations merit consideration. In real-world applications, calibration of the heterogeneity detection procedure relies on permutation-based approximations to the null distribution. While practical, such approaches are inherently constrained by the number of permutations and cannot guarantee exact Type I error control, in contrast to simulation settings where homogeneous scenarios can be explicitly specified. Developing more accurate and scalable inferential procedures for heterogeneity detection in data-driven subgroup analyses therefore represents an important direction for future research.

## Code Availability

An R package MLearner implementing the proposed method will be available on GitHub soon.

## Acknowledgments

The authors thank Amgen Inc. for supporting this research.

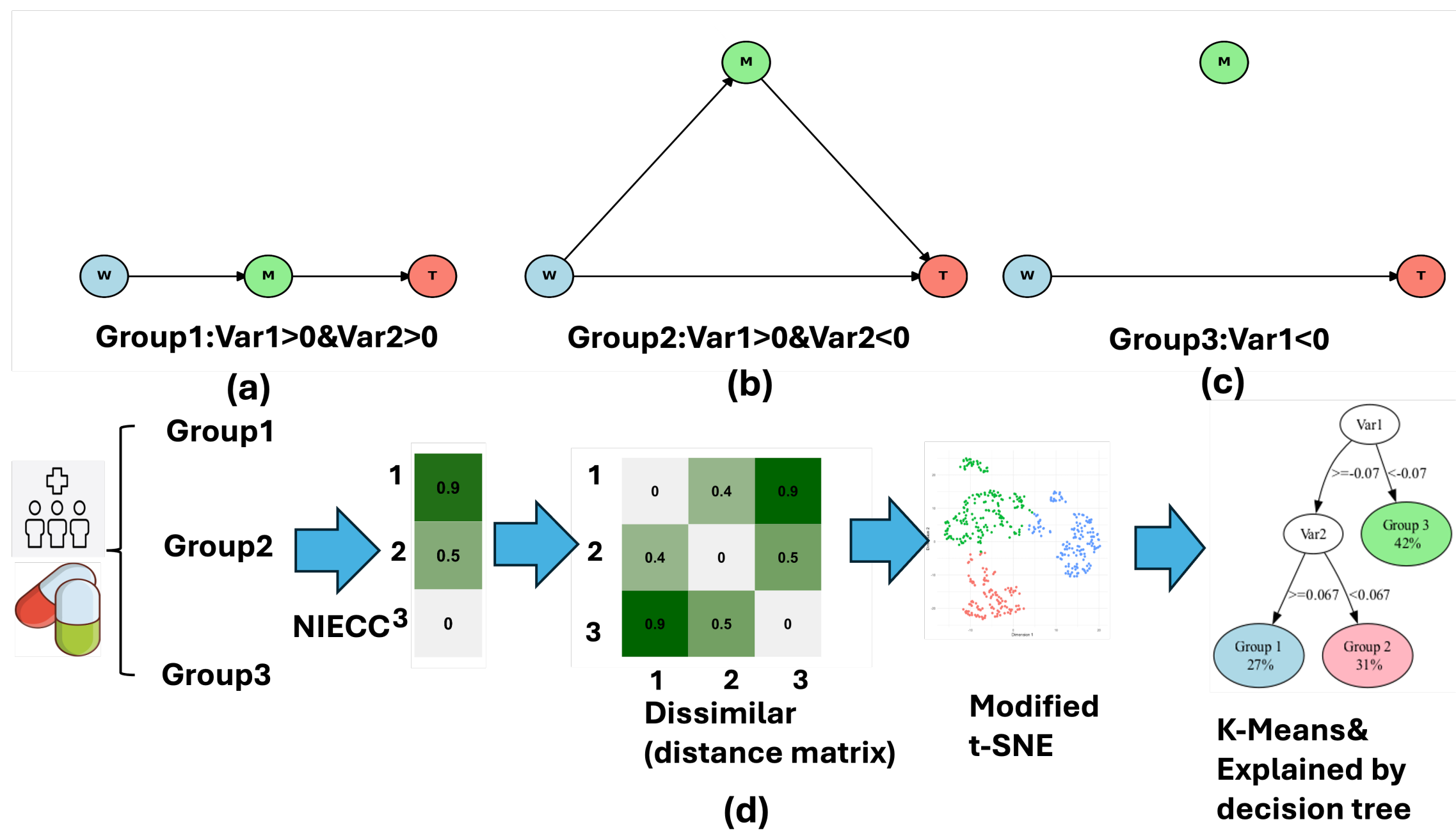

**Figure 1**: Causal diagrams for surrogate endpoint validation and mediation model. Panels (a)–(c) depict three scenarios of surrogate relationships. These scenarios provide illustrative examples for the analytical workflow presented in panel (d). (a) represents an ideal surrogate scenario, in which the surrogate endpoint $M$ fully captures the treatment effect on the clinical outcome $T$. (b) represents a non-ideal surrogate scenario, in which the surrogate endpoint captures only part of the treatment effect, with the remaining effect acting directly on the clinical outcome. $W$ denotes treatment, $M$ denotes surrogate biomarker, $T$ denotes clinical outcome. (c) represents $M$ is not surrogate endpoint, with no treatment effect mediated by $M$. (d) the overall pipeline of the proposed method. Illustration of subgroup identification when the surrogate is valid only in the region (Var1 > 0&Var2 > 0) and (Var1 > 0&Var2 < 0) and invalid elsewhere. The proposed method constructs a natural indirect treatment effect conditional on covariates (NIECC)-based dissimilarity matrix, projects observations into a two-dimensional space via a modified t-SNE, and then applies K-means clustering followed by a decision tree to obtain interpretable subgroup rules.

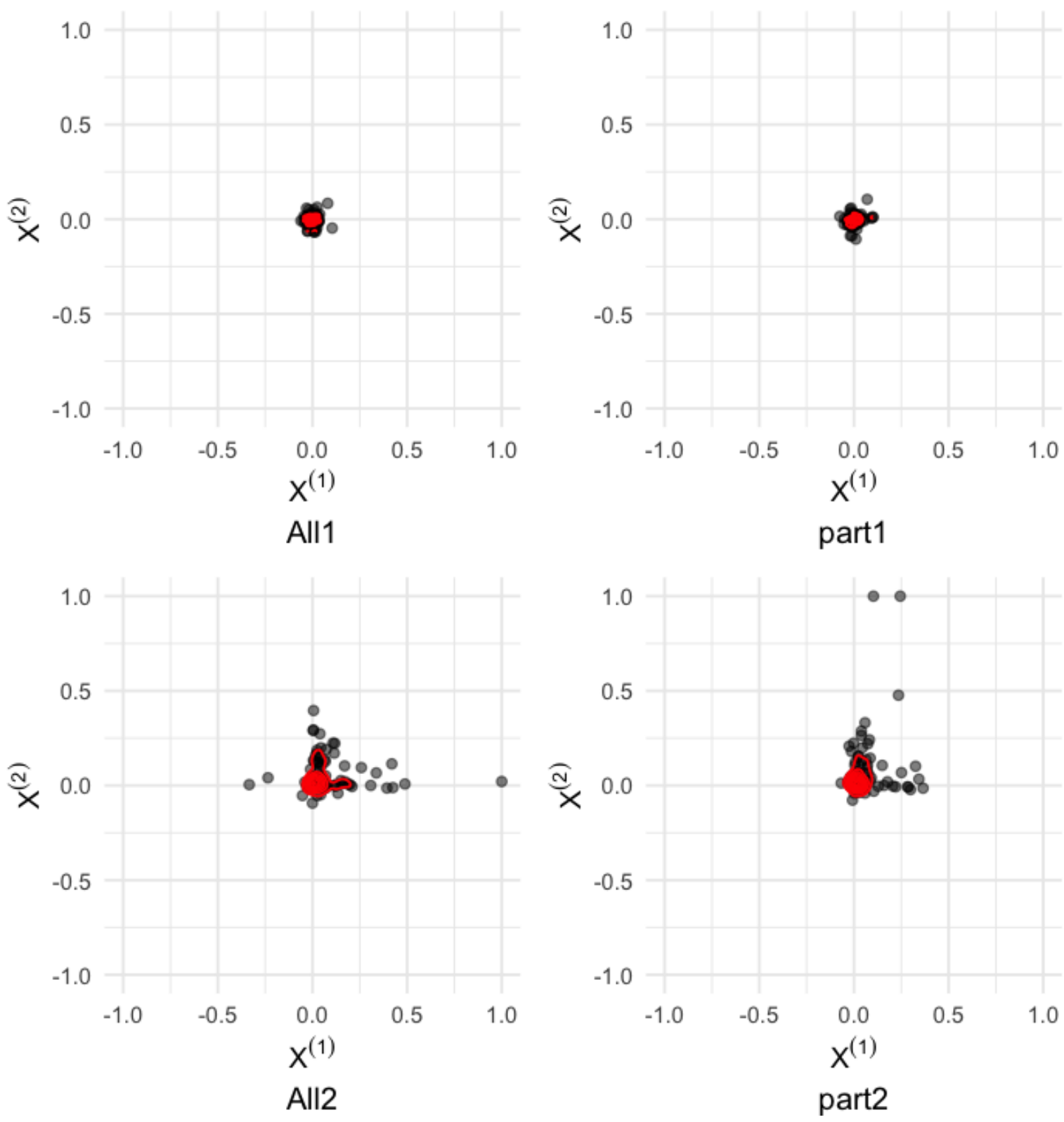

**Figure 2**: Distribution of thresholds of complex assumption. The mediator is active only in the region $X^{(1)} > 0$ and $X^{(2)} > 0$. Each point represents the estimated boundary threshold of the mediator-heterogeneous region obtained from the selected profile in a single simulation replicate. Black points denote replicate-specific estimates, and the red curves represent the corresponding density estimates. Scenarios All1 and All2 correspond to complete mediation of the treatment effect through the mediator, whereas Part1 and Part2 correspond to partial mediation. The treatment effect is constant in All1 and Part1, and covariate-dependent in All2 and Part2.

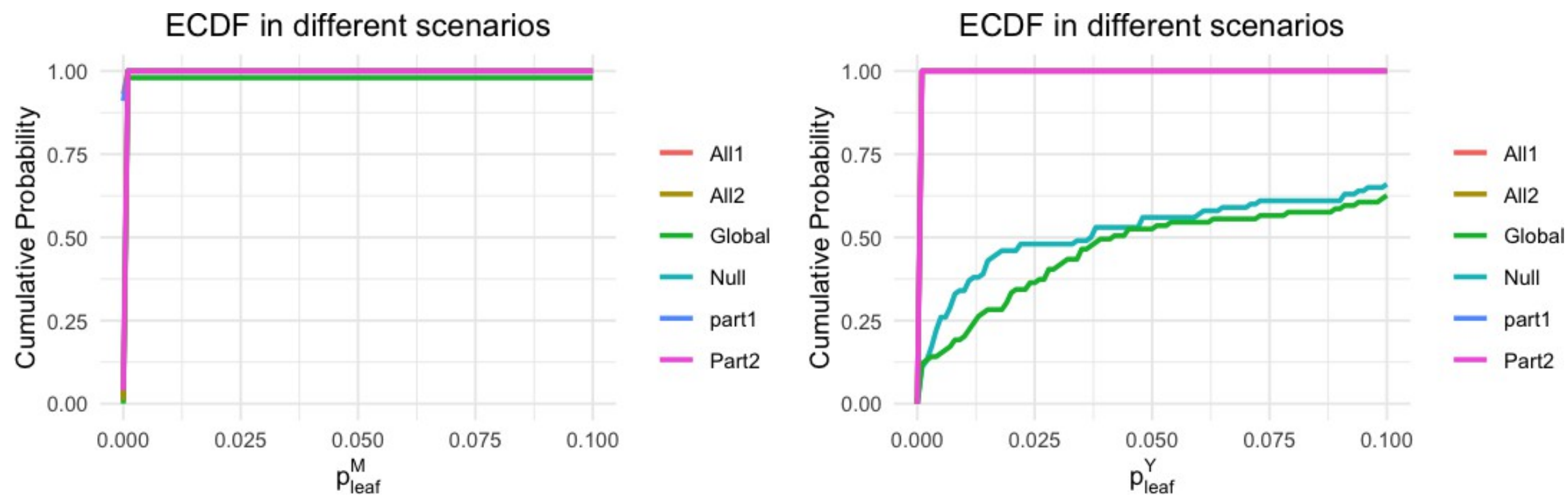

**Figure 3**: Empirical cumulative distribution function (ECDF) of $p^M_{leaf}$ and $p^Y_{leaf}$ in complex assumption. Scenarios All1 and All2 correspond to complete mediation of the treatment effect through the mediator, whereas Part1 and Part2 correspond to partial mediation. The treatment effect is constant in All1 and Part1, and covariate-dependent in All2 and Part2.

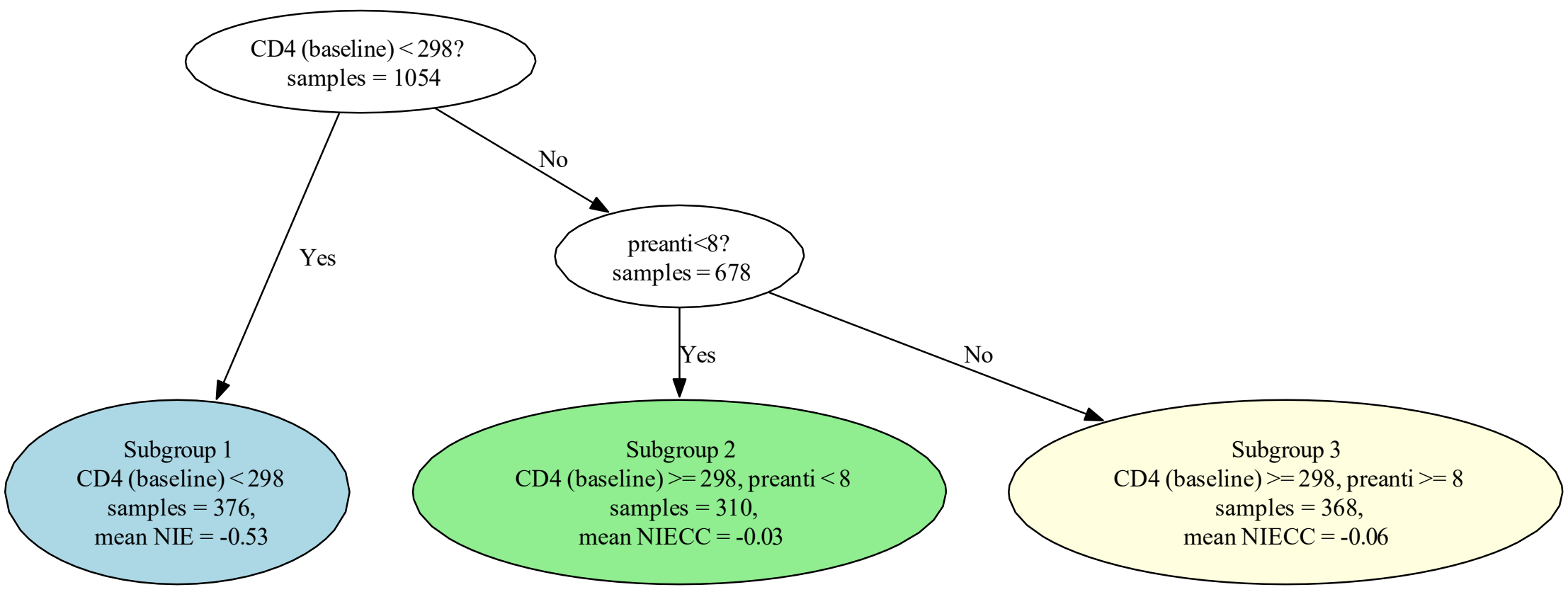

**Figure 4**: Profile of heterogeneous surrogate biomarker CD4 level at 20 weeks. mean NIECC: mean natural indirect treatment effect conditional on covariates in this group. preanti: prior antiretroviral therapy exposure measured as days of pre-ACTG 175 treatment. preanti < 8 days indicates that the duration of antiretroviral therapy is close to zero, corresponding to patients who are essentially treatment-na¨ıve.

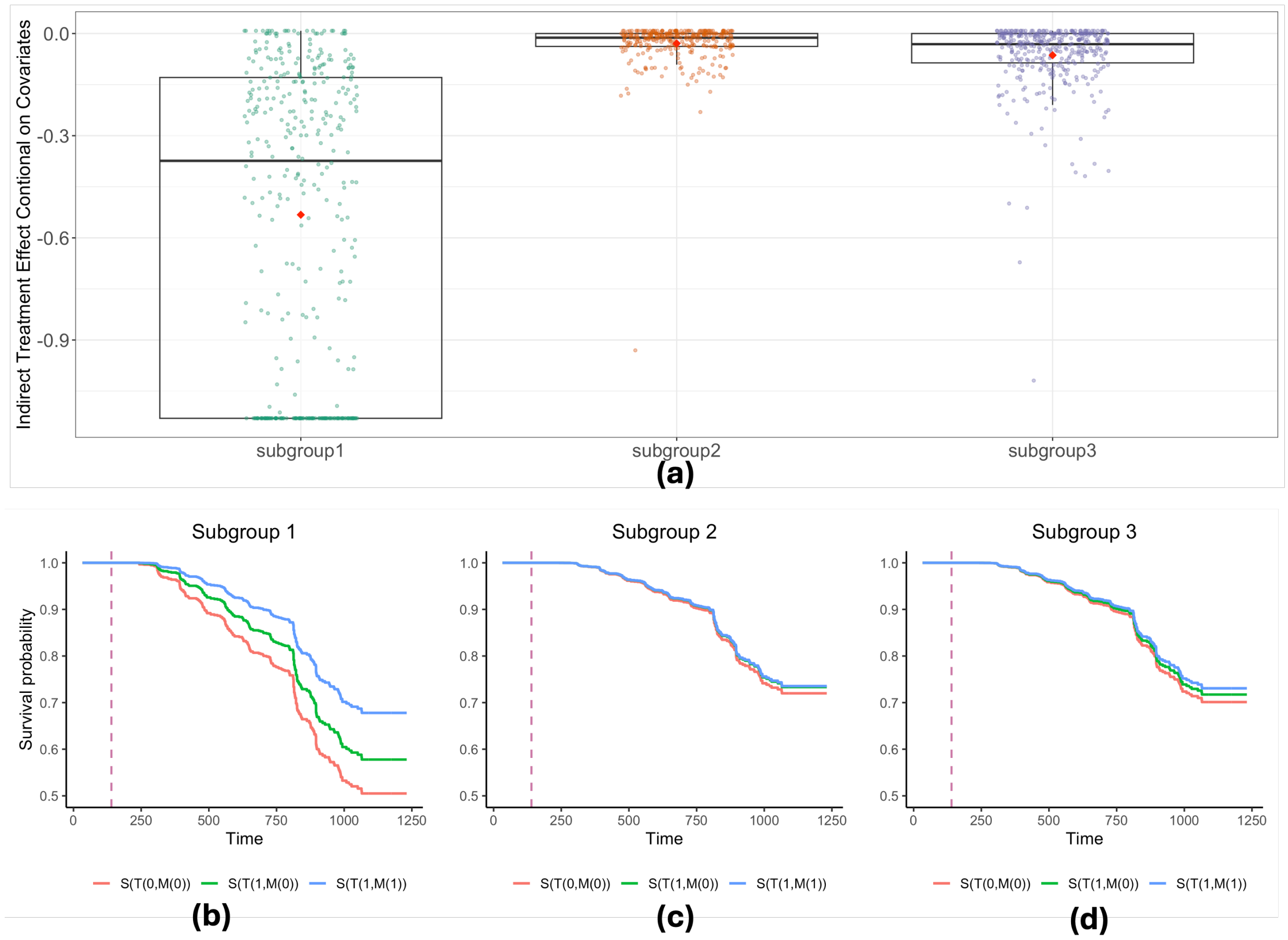

**Figure 5**: (a) Distribution of individual indirect treatment effects conditional on covariates across three subtypes. Each point represents an individual-level indirect treatment effect conditional on covariates. Boxes indicate the interquartile range with medians shown as horizontal lines, and red diamonds denote subgroup means. The mean indirect effects are −0.53 for Subgroup 1, −0.03 for Subgroup 2, and −0.06 for Subgroup 3. The reported values are intended to illustrate relative patterns and should not be interpreted as exact quantitative estimates. The indirect treatment effect estimates are truncated by displaying only individuals whose estimated indirect treatment effects lie between the 10th and 90th percentiles. The figures (b), (c), (d) show counterfactual survival curves under different treatment and mediator settings. The curve $S(1, M(1))$ represents the survival probability if all individuals received treatment and the mediator took its natural value under treatment. The curve $S(1, M(0))$ represents the survival probability if all individuals received treatment while the mediator was fixed at the level it would have taken under control. The curve $S(0, M(0))$ represents the survival probability if all individuals received control and the mediator took its natural value under control. The vertical purple line indicates the time of surrogate measurement (20 weeks, i.e., 140 days). (a) Subgroup 1, (b) Subgroup 2, (c) Subgroup 3. Details on the construction of the counterfactual survival plots are provided in Appendix I.

# Supporting Information for Modeling Heterogeneous Mediation Effects in Survival Analysis via an Interpretable M-Learner Framework by Xingyu Li, Qing Liu, Xun Jiang, Hong Amy Xia, Brian P. Hobbs, and Peng Wei

# 1 Web Appendix A: Heterogeneous mediation treatment effects

The outcome model is specified as

$$\lambda(t|\boldsymbol{X}, W, M) = \lambda_0(t) exp(g(\boldsymbol{X}, W, M)), \quad (1)$$

where

$$g(\boldsymbol{X}, W, M) = \kappa_1(\boldsymbol{X}) + \kappa_2(\boldsymbol{X}) \cdot W + \kappa_3(\boldsymbol{X}) \cdot M. \quad (2)$$

The mediator is modeled as

$$M = \kappa_4(\boldsymbol{X}) + \kappa_5(\boldsymbol{X}) \cdot W + \epsilon, \quad (3)$$

where $\kappa_4(\boldsymbol{X})$ represents the baseline covariate effects on the mediator, $\kappa_5(\boldsymbol{X})$ denotes the treatment effect on the mediator, and $\epsilon$ is a mean-zero term.

$$\begin{aligned} ITE(t|\boldsymbol{X} &= \boldsymbol{x}) = \kappa_3(\boldsymbol{x})\kappa_5(\boldsymbol{x}) = g(\boldsymbol{x}|W = 1, M = \mathrm{E}[M|w = 1, \boldsymbol{X} = \boldsymbol{x}]) - \\ & g(\boldsymbol{x}|W = 1, M = \mathrm{E}[M|w = 0, \boldsymbol{X} = \boldsymbol{x}]). \end{aligned} \quad (4)$$

Equation (4) represents the NIECC. We now discuss its implications.

If $\kappa_1(\boldsymbol{x})$, $\kappa_2(\boldsymbol{x})$, and $\kappa_4(\boldsymbol{x})$ are specified as linear functions of $\boldsymbol{x}$, and both $\kappa_3(\boldsymbol{x})$ and $\kappa_5(\boldsymbol{x})$ are constants, that is, $\kappa_3(\boldsymbol{x}) = \beta_m$ and $\kappa_5(\boldsymbol{x}) = \beta_{wm}$, then the NIECC simplifies to

$$ITE(t \mid \boldsymbol{X} = \boldsymbol{x}) = \beta_w \beta_{wm},$$

which coincides with the classical mediation result [VanderWeele, 2011].

However, under this specification, the indirect treatment effect does not vary with covariates. Consequently, the model cannot capture heterogeneity in the indirect effect across individuals with different covariates profiles.

To capture the variation of the indirect treatment effect with respect to covariates, we consider two scenarios: a simplified linear specification and a more flexible, complex specification. We discuss each case separately in the following sections.

## 1.1 Linear assumption

We first consider a simplified linear specification in which $\kappa_1(\boldsymbol{X})$ and $\kappa_2(\boldsymbol{X})$ are assumed to be linear functions of $\boldsymbol{X}$. If $\kappa_3(\boldsymbol{X})$ is also specified as linear in $\boldsymbol{X}$, the hazard model would involve a large number of interaction terms of the form $M \times \boldsymbol{X}$, leading to substantial model complexity when the number of covariates is moderate or high. Such an expansion increases the dimensionality of the parameter space and may result in unstable estimation.

To maintain parsimony while preserving flexibility in modeling heterogeneous mediation effects, we instead assume a constant effect of the mediator on the hazard, that is, $\kappa_3(\boldsymbol{X}) =$ $\beta_m$, while allowing $\kappa_5(\boldsymbol{X})$ to follow a flexible nonlinear specification. Under this structure, the heterogeneity of the indirect treatment effect is fully determined by $\kappa_5(\boldsymbol{X})$, yielding

$$ITE(t \mid \boldsymbol{X} = \boldsymbol{x}) = \beta_m \, \kappa_5(\boldsymbol{x}),$$

which allows the effect to vary with covariates while avoiding excessive parameterization in the hazard model.

So in this setting,

$$\begin{aligned} ITE(t \mid \boldsymbol{X} = \boldsymbol{x}) &= \kappa_3(\boldsymbol{x})\kappa_5(\boldsymbol{x}) \\ &= \beta_m \, \kappa_5(\boldsymbol{x}) \\ &= \beta_m \left\{ \mathrm{E}[M \mid W = 1, \boldsymbol{X} = \boldsymbol{x}] - \mathrm{E}[M \mid W = 0, \boldsymbol{X} = \boldsymbol{x}] \right\}. \end{aligned} \tag{5}$$

In this framework, the coefficient $\beta_m$ can be estimated using standard parametric methods within a Cox proportional hazards regression model. In contrast, the conditional expectations $\mathrm{E}[M \mid W = 1, \boldsymbol{X} = \boldsymbol{x}]$ and $\mathrm{E}[M \mid W = 0, \boldsymbol{X} = \boldsymbol{x}]$ can be estimated using flexible regression or machine learning methods. This hybrid strategy enables flexible modeling of heterogeneous mediation effects while retaining a parsimonious hazard specification.

The problem of regularized kernel learning covers a broad class of methods that have been thoroughly studied in the statistical learning literature ([Bartlett and Mendelson, 2006, Steinwart and Christmann, 2008, Steinwart et al., 2009, Mendelson and Neeman, 2010, Nie and Wager, 2021]), and thus provides an ideal case study for examining the property of $\mathrm{E}[M \mid W = 1, \boldsymbol{X} = \boldsymbol{x}]$ or $\mathrm{E}[M \mid W = 0, \boldsymbol{X} = \boldsymbol{x}]$.

We study $\|\cdot\|_{\mathcal{H}}$-penalized kernel regression, where $\mathcal{H}$ is a reproducing kernel Hilbert space (RKHS) with a continuous, positive semidefinite kernel function $K$.

Let $\mathcal{X} \subset \mathbb{R}^d$ be a compact metric space and let $P$ denote the distribution of $X$. Let $K : \mathcal{X} \times \mathcal{X} \to \mathbb{R}$ be a continuous positive semidefinite kernel and let $\mathcal{H}$ be the associated RKHS.

Define the integral operator $T_K : L_2(P) \to L_2(P)$ by

$$T_K(f)(\cdot) = \mathrm{E}\left[ K(\cdot, X) f(X) \right].$$

By Mercer's theorem [Cucker and Smale, 2002], there exists an orthonormal basis of eigenfunctions $\{\psi_j\}_{j\geq 1}$ in $L_2(P)$ with corresponding eigenvalues $\{\sigma_j\}_{j\geq 1}$ such that

$$K(x, y) = \sum_{j=1}^{\infty} \sigma_j \psi_j(x) \psi_j(y).$$

Define the feature map $\phi : \mathcal{X} \to \ell_2$ by

$$\phi(x) = \left( \sqrt{\sigma_j} \psi_j(x) \right)_{j=1}^{\infty}.$$

Following Mendelson and Neeman [2010], the RKHS $\mathcal{H}$ can be identified with the image of $\ell_2$, where for every $t \in \ell_2$ the corresponding function in $\mathcal{H}$ is

$$f_t(x) = \langle \phi(x), t \rangle,$$

with inner product

$$\langle f_s, f_t \rangle_{\mathcal{H}} = \langle s, t \rangle_{\ell_2}.$$

We estimate the mediator regression function

$$\phi_{M_1}(\boldsymbol{x}) = \mathrm{E}[M \mid \boldsymbol{X} = \boldsymbol{x}, W = 1]$$

using a regularized kernel estimator

$$\hat{\phi}_{M_1} = \arg\min_{f \in \mathcal{H}} \left( \frac{1}{n_1} \sum_{i:W_i=1} (M_i - f(\boldsymbol{X}_i))^2 + \lambda \|f\|_{\mathcal{H}}^2 \right),$$

where $n_1$ denotes the number of treated observations and $\lambda > 0$ is a regularization parameter.

**Assumption S1.** Without loss of generality, assume $K(\boldsymbol{x}, \boldsymbol{x}) \leq 1$ for all $\boldsymbol{x} \in \mathcal{X}$. For some $0 < p < 1$, the eigenvalues $\sigma_j$ satisfy

$$G = \sup_{j \geq 1} j^{1/p} \sigma_j < \infty,$$

and the eigenfunctions $\psi_j$ with $\|\psi_j\|_{L_2(P)} = 1$ are uniformly bounded,

$$\sup_j \|\psi_j\|_\infty \leq A < \infty.$$

Finally, the mediator satisfies

$$|M| \leq C_M \quad \text{almost surely.}$$

**Assumption S2.** The true mediator function

$$\phi_{M_1}(\boldsymbol{x}) = \mathrm{E}[M \mid \boldsymbol{X} = \boldsymbol{x}, W = 1]$$

satisfies

$$T_K^{-\alpha} \phi_{M_1} \in L_2(P)$$

for some $0 < \alpha < 1$, i.e.

$$\mathrm{E}\,(T_K^{-\alpha} \phi_{M_1}(\boldsymbol{X}))^2 < \infty.$$

Assumptions S1–S2 are standard regularity conditions for kernel-based estimators and coincide with those used in [Mendelson and Neeman, 2010]. Assumption S1 controls the complexity of the RKHS through the eigenvalue decay of $T_K$, while Assumption S2 is a source condition that characterizes the smoothness of $M_1$ relative to the kernel.

*Theorem S1.* Under Assumptions S1–S2 and for a suitable choice of the regularization parameter $\lambda$, the kernel estimator satisfies

$$R(\hat{\phi}_{M_1}) = \tilde{O}_P\left(n^{-\frac{2\alpha}{p+2\alpha}}\right),$$

where

$$R(f) = \mathrm{E}\left[(f(X) - \phi_{M_1}(X))^2\right]$$

and $\tilde{O}_P(\cdot)$ hides logarithmic factors.

In particular, $\|\hat{\phi}_{M_1} - \phi_{M_1}\|^2_{L_2(P)} = \tilde{O}_P\left(n^{-\frac{2\alpha}{p+2\alpha}}\right)$, which the rate was established in [Mendelson and Neeman, 2010] and the suitable choice of the regularization parameter $\lambda$ could be found at [Mendelson and Neeman, 2010].

The estimation error admits the following bound:

$$\begin{aligned} \|\hat{\beta}_m \hat{\phi}_{M_1} - \beta_m \phi_{M_1}\|_{L_2(P)} &\leq \|\phi_{M_1}\| |\hat{\beta}_m - \beta_m| + |a| \|\hat{\phi}_{M_1} - \phi_{M_1}\| \\ &= O_p(1) \cdot O_p(n^{-1/2}) + O_p(1) \cdot O_p(n^{-\frac{\alpha}{p+2\alpha}}) \\ &= O_p(n^{-\frac{\alpha}{p+2\alpha}}) \end{aligned} \tag{6}$$

This result confirms that the overall convergence rate is dictated by the non-parametric component $\hat{\phi}$, independent of whether sample splitting is implemented.

### 1.1.1 Estimation

Under the linear assumption, we propose an estimation procedure corresponding to the decomposition

$$ITE(t \mid \mathbf{X} = \mathbf{x}) = \beta_m \{\mathrm{E}[M \mid W = 1, \mathbf{X} = \mathbf{x}] - \mathrm{E}[M \mid W = 0, \mathbf{X} = \mathbf{x}]\}.$$

Specifically, we first estimate the coefficient $\beta_m$ using a parametric model for the outcome. We then estimate the conditional expectation of the mediator given covariates separately within the treatment and control arms. We denote the estimator of $\mathrm{E}[M \mid W = 1, \mathbf{X} = \mathbf{x}]$ by $\hat{m}_1(\boldsymbol{x})$ and the estimator of $\mathrm{E}[M \mid W = 0, \mathbf{X} = \mathbf{x}]$ by $\hat{m}_0(\mathbf{x})$. The plug-in estimator of the NIECC is obtained by combining these estimates:

$$\widehat{ITE}(t \mid \mathbf{X} = \mathbf{x}) = \hat{\beta}_m \{\hat{m}_1(\mathbf{x}) - \hat{m}_0(\mathbf{x})\}.$$

## 1.2 Complex assumption

Under the linear assumption, heterogeneity in NIECC can only be captured through the mediator $M$. However, in practice the mediator itself may exhibit treatment-induced heterogeneity, and the clinical outcome may respond differently across levels of $M$. In other words, the effect of $M$ (i.e., its coefficient) may also vary with covariates and thus act as an effect modifier.

For example, in HIV studies, the impact of CD4 count on survival may differ substantially across patients with different CD4 levels. Moreover, changes in CD4 count are influenced not only by treatment but also by individual covariate profiles. Therefore, both the mediator process and the outcome response to the mediator may exhibit substantial heterogeneity, which cannot be adequately represented under a purely linear specification.

Therefore, the $\kappa_3(\boldsymbol{X})$ could be nonlinear function. However, it is impossible to identify $\kappa_3(\boldsymbol{X})$, so we identify the NIECC by

$$
\begin{aligned}
ITE(t|\boldsymbol{X}=\boldsymbol{x}) &= \kappa_3(\boldsymbol{x})\kappa_5(\boldsymbol{x}) \\
&= g(\boldsymbol{x}|W=1, M=\mathrm{E}[M|w=1, \boldsymbol{X}=\boldsymbol{x}]) - \\
&\quad\; g(\boldsymbol{x}|W=1, M=\mathrm{E}[M|w=0, \boldsymbol{X}=\boldsymbol{x}]). \qquad (7)
\end{aligned}
$$

### 1.2.1 Estimation

When estimating the NIECC under the complex assumption based on (7), the estimation of the conditional expectation of the mediator remains the same as that under the linear assumption. For the estimation of $g(\boldsymbol{x}|\ W=1, M=m)$, we adopt a flexible modeling strategy that allows for nonlinear relationships and interactions between the mediator and covariates.

We consider use nonparametric method based on Cox loss plus regularization (the loss function is defined as the negative partial log-likelihood (8) plus regularization) to estimate $g$, such as XGBoost and deep learning method (Deepsurv)[Chen and Guestrin, 2016, Katzman et al., 2018, Liu et al., 2020].

We consider the Cox proportional hazards model

$$\lambda(t \mid \boldsymbol{Z}) = \lambda_0(t)\exp(g(\boldsymbol{Z})),$$

where $\boldsymbol{Z} = (\boldsymbol{X}, W, M)$.

Define the risk set at time $U_i$ as

$$R(U_i) = \{j : U_j \geq U_i\}.$$

The Cox partial likelihood is defined as

$$L(g) = \prod_{i:\delta_i=1} \frac{\exp(g(\boldsymbol{Z}_i))}{\sum_{j\in R(U_i)} \exp(g(\boldsymbol{Z}_j))}.$$

The empirical Cox loss is defined as

$$\ell_n(g) = -\frac{1}{n}\sum_{i=1}^{n} \delta_i \left( g(\boldsymbol{Z}_i) - \log \sum_{j\in R(U_i)} \exp(g(\boldsymbol{Z}_j)) \right). \qquad (8)$$

Denote

$$\hat{g} = \arg\min_{g} \ell_n(g).$$

Due to the invariance property of the Cox loss, the estimator $\hat{g}$ is identifiable only up to an additive constant. In particular, there exists a constant $c$ such that

$$\hat{g}(\boldsymbol{z}) \approx g^{*}(\boldsymbol{z}) + c.$$

However, when estimating the NIECC defined through contrasts of $g$, the additive constant cancels out. Hence the NIECC is invariant to this non-identifiability and can be consistently estimated. Denote

$$\hat{g} = \arg\min_{g} \ell_n(g).$$

Because the Cox loss determines $g$ only up to an additive constant, the estimator $\hat{g}$ may differ from the population optimal function $g^{*}$ by a constant shift, i.e.,

$$\hat{g}(\boldsymbol{z}) \approx g^{*}(\boldsymbol{z}) + c.$$

Because the function $g$ is estimated using data from the treatment arm ($W = 1$), the estimator of the NIECC is constructed based on contrasts of the predicted outcome under different mediator levels. In particular, the NIECC is defined through differences in $g$, for example

$$g(\boldsymbol{x} \mid W = 1, M = \mathrm{E}[M \mid \boldsymbol{X} = \boldsymbol{x}, W = 1]) - g(\boldsymbol{x} \mid W = 1, M = \mathrm{E}[M \mid \boldsymbol{X} = \boldsymbol{x}, W = 0]).$$

Although the function $\hat{g}$ may be identifiable only up to an additive constant when estimated from the treatment arm data, this constant cancels out when forming such contrasts. Therefore, the constant non-identifiability of $\hat{g}$ does not affect the estimation of the NIECC.

# 2 Web Appendix B: Clustering

Recent advances in biomedical and digital health technologies have enabled increasingly granular profiling of patient characteristics, facilitating the identification of distinct disease subgroups and the evaluation of drug-specific prognostic effects. Although patients may not be fully statistically exchangeable, individuals who are sufficiently similar with respect to key clinical features can be leveraged for prognostic modeling and subgroup discovery. Accordingly, precision medicine requires methodologies that quantify patient similarity along a continuum rather than relying on strict partitioning assumptions [Sharafoddini et al., 2017, Parimbelli et al., 2018]. Similarity-based approaches have shown considerable promise in improving comparative effectiveness research and personalized prediction, particularly in the presence of heterogeneity. For example, Lee et al. [2015] demonstrated that, under heterogeneous treatment effects, predictive performance can be improved by focusing on a subset of patients who are most similar to a target individual rather than using the entire cohort. In the context of surrogate evaluation and subgroup discovery, similarity should be assessed with respect to patient characteristics and treatment assignment, while being supervised by expected clinical outcomes (such as OS) and surrogate behavior (such as tumor response or CD4 count). Importantly, different covariates may contribute unequally to outcome prediction and surrogate relevance, motivating the need to weight patient profile variables according to their importance. This can be achieved through supervised similarity learning, in which outcome and surrogate information guide the construction of patient similarity in the covariate space. Empirical evidence supports the advantages of this strategy; for instance, in a heart failure treatment recommendation task using electronic health records, Panahiazar et al. [2015] showed that supervised clustering substantially outperforms unsupervised methods such as K-means and hierarchical clustering. Formally, consider a supervised learning framework with $n$ training samples $\{X_1, \ldots, X_n\}$, each characterized by $p$ baseline covariates, corresponding outcomes $\{Y_1, \ldots, Y_n\}$, and surrogate biomarkers $\{M_1, \ldots, M_n\}$. In our framework, the surrogate is not treated as an independent input, but instead is modeled as a function of covariates and treatment, thereby aligning with the framework of indirect treatment effects conditional on covariates. Consequently, both outcome and surrogate information are used solely to supervise the construction of similarity among covariate profiles. Specifically, we learn a prediction function $h : \mathcal{X} \to \mathcal{Y}$ by minimizing

$$\sum_{i=1}^{n} L\{h(X_i, M(X_i)), Y_i\},$$

while jointly modeling the conditional distribution of the surrogate $M$ given $X$. Here, $M$ is treated as a function of $X$, rather than an independent input, capturing the surrogate mechanism conditional on covariates. Based on the predicted outcomes and induced surrogate responses, we construct an embedding

$$\phi(\cdot) : \mathcal{X} \to \mathcal{S},$$

whose geometry reflects similarity in predicted responses rather than raw covariate space.

The induced similarity structure is defined as $S(i, j) = \phi\{h(X_i), h(X_j)\}$, such that nearby points in the embedding correspond to patients with similar predicted clinical outcomes and similar conditional surrogate behavior. This embedding captures heterogeneity in indirect treatment effects through covariates, yielding a representation that reflects similarity in both patient characteristics and predicted response patterns, while maintaining a principled mapping from covariates to similarity.

From a clinical trial perspective, the proposed clustering strategy facilitates the identification of patient subpopulations for whom the surrogate biomarker captures the treatment effect in a consistent and interpretable manner, thereby enabling refined subgroup analyses and more reliable treatment evaluation. Given a dissimilarity (distance) matrix, a wide range of clustering strategies may be applied. Many existing approaches first transform the dissimilarity matrix into a similarity matrix and subsequently employ clustering techniques tailored to similarity-based representations, such as spectral clustering or community detection [Li et al., 2026]. These methods, however, are inherently designed for similarity matrices and often rely on additional structural assumptions on the induced similarity graph.

In contrast, we adopt an alternative strategy that directly operates on the dissimilarity matrix by applying a modified t-distributed stochastic neighbor embedding (t-SNE) procedure to embed patients into a low-dimensional Euclidean space. This embedding enables flexible clustering using standard algorithms such as K-means, while preserving heterogeneity in surrogate-mediated treatment effects. t-SNE is a widely used technique for visualizing high-dimensional data, which operates by quantifying pairwise relationships between observations and constructing a low-dimensional embedding that preserves local neighborhood structure [Van der Maaten and Hinton, 2008]. Specifically, t-SNE seeks an embedding that minimizes the Kullback–Leibler divergence between the pairwise similarity distribution induced by the original dissimilarities and the corresponding similarity distribution in the low-dimensional space.

In our framework, distances are not computed in the original feature space but are instead defined through the NIECC. Consequently, the resulting low-dimensional embedding captures similarity among patients in terms of their NIECC, rather than their observed covariates. Clustering is then performed on this embedding using the K-means algorithm to identify subgroups of patients with homogeneous indirect treatment effects.

To facilitate interpretation, we further consider a two-dimensional representation of this embedding as a visualization of the underlying clustering structure. At the outset, this raises fundamental questions regarding what constitutes a "good" visualization of clustering. These challenges parallel well-known difficulties in formalizing clustering objectives more broadly [Kleinberg, 2002]. In theoretical analyses, such issues are often addressed by adopting a standard clustering formalization and assuming that the data are generated from an unknown ground-truth clustering structure [Kleinberg, 2002, Arora et al., 2018]. Following this perspective, we formalize the visualization objective as constructing a two-dimensional embedding in which points within the same cluster are substantially closer to one another than to points in different clusters.

Arora et al. [2018] analyzed the theoretical properties of the original t-SNE under a regime in which the number of clusters is allowed to grow with the sample size $n$, our analysis focuses on a fundamentally different setting motivated by clinical applications. Specifically, we study a modified t-SNE procedure in which pairwise dissimilarities are defined based on similarities in NIECC, rather than on raw feature distances. Moreover, unlike the growing-cluster regime considered in [Arora et al., 2018], we assume that the number of clinically meaningful subgroups is finite and does not increase with $n$. This fixed-subgroup assumption is well aligned with clinical practice, where interpretability and clinical relevance constrain the number of actionable patient subpopulations. Our theoretical results therefore characterize the behavior of modified t-SNE under a regime that is both methodologically and substantively distinct from existing analyses.

Under mild regularity conditions, we show that t-SNE provably produces meaningful visualizations under this finite-cluster regime. We first formalize the visualization problem and then establish theoretical guarantees for visualizing heterogeneity in NIECC. In addition, we provide the provable guarantees for a modified t-SNE algorithm applied to clusterable data under a fixed number of clusters.

**Formalizing Visualization.** We assume that the NIECC, denoted by $v_1, \cdots, v_n \in \mathrm{V}$, and that there exists an underlying "ground-truth" clustering represented by a partition $\mathrm{C}_1, \cdots, \mathrm{C}_k$ of the index set $[n]$ into $k$ clusters, where $[n] := \{1, 2, \ldots, n\}$.

A visualization is given by a 2-dimensional embedding $Y = \{y_1, \cdots, y_n\} \subseteq \mathbb{R}^2$ of $V$, where each $v_i \in V$ is mapped to the corresponding $y_i \in Y$. Intuitively, a cluster $C_l$ in the original data is visualized if the corresponding points in the 2-dimensional embedding $Y$ are well-separated from all the rest. The following definitions from formalizes this idea [Arora et al., 2018].

**Definition 1** (Visible cluster)**.** Let $Y$ be a 2-dimensional embedding of a dataset $V$ with ground-truth clustering $C_1, \cdots, C_k$. Given $\epsilon \geq 0$, a cluster $C_l$ in $V$ is said to be $(1 - \epsilon)$-visible in $Y$ if there exist $P, P_{err} \subseteq [n]$ such that:

1. $|(P \setminus C_l|) \cup (C_l \setminus P)| \leq \epsilon|C_l|$, $P_{err} \leq \epsilon n$, and
2. for each $i, i' \in P$ and $j \in [n] \setminus (P \cup P_{err})$, $\|y_i - y_{i'}\| \leq \frac{1}{2}\|y_i - y_j\|$.

In this setting, we say that $P(1 - \epsilon)$- visualizes $C_i$ in $Y$.

We define $Y$ to be a good visualization if every cluster $C_\ell$ in the dataset $V$ is faithfully represented in $Y$.

**Definition 2** (Visualization)**.** Let $Y$ be a 2-dimensional embedding of a dataset $V$ with ground-truth clustering $C_1, \cdots, C_k$. Given $\epsilon \geq 0$, we say that $Y$ is a $(1 - \epsilon)$-visualization of $V$ if there exists a partition $P_1, \cdots, P_k, P_{err}$ of $[n]$ such that:

1. For each $i \in [k]$, $P_i(1 - \epsilon)$-visualizes $C_i$ in $Y$, and
2. $P_{err} \leq \epsilon n$.

In particular, when $\epsilon = 0$, we say that $Y$ is a full visualizationi of $V$.

*Remark* 1. A natural question is whether the clustering inferred from a visualization is unique. Our definition above does not guarantee uniqueness, and indeed, such a guarantee is inherently impossible due to the intrinsic ambiguity in the notion of clustering. For instance, it may be unclear whether a given set of points should be treated as a single cluster or as two smaller, distinct clusters.

Nevertheless, under an additional assumption that the size of any cluster is less than twice that of any other cluster, it can be formally established that a full visualization, as defined in Definition 2, uniquely determines the clustering.

**Definition 3** (Well-separated, spherical data)**.** Let $V = \{v_1, \cdots, v_n\}$ be clusterable data with $C_1, \cdots, C_k$ defining the individual clusters such that for each $l \in [k]$, $|C_l| \geq 0.1(n/k)$. We say that $V$ is $\gamma-$spherical and $\gamma-$well-separated if for some $b_1, \cdots, b_k > 0$, we have:

1. $\gamma$-spherical: For any $l \in [k]$ and $i, j \in C_l (i \neq j)$, we have $(v_i - v_j)^2 \geq \frac{b_l}{1+\gamma}$, and for any $i \in C_l$ we have $|\{j \in C_l \setminus \{i\} : (v_i - v_j)^2 \leq b_l\}| \geq 0.51|C_l|$.
2. $\gamma$-well-sparated: For any $l, l' \in [k] (l \neq l')$, $i \in C_l$ and $j \in C_{l'}$ we have $(v_i - v_j)^2 \geq (1 + \gamma \log n) \max\{b_l, b_{l'}\}$.

The first condition asks for the distances between points in the same cluster ("intra-cluster distances") to be concentrated around a single value (with controlling the "amount" of concentration). The second condition requires that the distances between two points from different clusters should be somewhat larger than the intra-cluster distances for each of the two clusters involved. In addition, we require that none of the clusters has too few points. These assumptioins have been used in previous work studing "distance-based" clustering algorithms [Dasgupta, 1999, Arora and Kannan, 2005, Arora et al., 2018].

**The modified t-SNE Algorithm.** The modified t-SNE algorithm (hereafter referred to as t-SNE) starts by computing a joint probability distribution $p_{ij}$ over pairs of points $v_i$, $v_j$, $i \neq j$ :

$$p_{j|i} = \frac{\exp\left(-(v_i - v_j)^2/2\tau_i^2\right)}{\sum_{l\in[n]\setminus\{i\}} \exp\left(-(v_i - v_l)^2/2\tau_i^2\right)}, \quad p_{ij} = \frac{p_{i|j} + p_{j|i}}{2n}, \tag{9}$$

where $\tau_i$ is a tunable parameter that controls the bandwidth of the Gaussian kernel around point $v_i$. In a two-dimensional map $\mathrm{Y} = \{y_1, \cdots, y_n\} \subset \mathrm{R}^2$, define the affinity $q_{ij}$ between points $y_i$ and $y_j$ ($i \neq j$) as

$$q_{ij} = \frac{(1 + \|y_i - y_j\|^2)^{-1}}{\sum_{l,s\in[n], l\neq s}(1 + \|y_l - y_s\|^2)^{-1}}. \tag{10}$$

The t-SNE tries to find points $y_i$'s in $\mathrm{R}^2$ that minimize the KL-divergence between $p$ and $q$:

$$f(y_1, \cdots, y_n) := KL(p||q) = \sum_{i,j\in[n], i\neq j} p_{ij} \log \frac{p_{ij}}{q_{ij}}. \tag{11}$$

The objective function $f$ is minimized using gradient descent. Its gradient is the following:

$$\frac{\partial f}{\partial y_i} = 4 \sum_{j\in[n]\setminus\{i\}} (p_{ij} - q_{ij}) q_{ij} Z (y_i - y_j), \quad i \in [n],$$

where $Z = \sum_{l,s\in[n], l\neq s}(1 + \|y_l - y_s\|^2)^{-1}$.

The t-SNE are based on an early exaggeration stage followed by an embedding stage that iterates a certain gradient descent algorithm. In early exaggeration stage, for the first $T$ iterations, the t-SNE select a larger exaggeration parameter $\alpha$, then in embedding stage, t-SNE drops and the original iterative Algorithm 1 is carried out till attaining a prespecified number of steps. In particular, it is empirically observed that, the early exaggeration technique enables t-SNE to find a better global structure, in the early stages of the optimization by creating very tight clusters of points that easily move around in the embedding space [Van Der Maaten, 2014]. Cai and Ma [2022] supports the phenomenon by theoretical analysis. For spherical and well-separated data, our theorem below shows that t-SNE with early exaggeration succeeds in finding a full visualization. In the early exaggeration, all $p_{ij}$'s are multipled by a factor $\alpha > 1$ [Van der Maaten and Hinton, 2008]. Letting the step size in the gradient descent method be $\frac{h}{4}$, we get the following update rule:

$$y_{i,(t+1)} = y_{i,(t)} + h \sum_{j\in[n]\setminus\{i\}} (\alpha p_{ij} - q_{ij,(t)}) q_{ij,(t)} Z_{(t)} (y_{j,(t)} - y_{i,(t)}), \; i = 1, \cdots, n. \tag{12}$$

Here, $y_{i,(t)} \in \mathbb{R}^2$ is the position of $y_i$ after $t$ iterations. We summarize the t-SNE algorithm in Algorithm 1, we let $\{y_{i,(0)}\}$ be initialized i.i.d from the uniform distribution over $[-0.01, 0.01]^2$.

*Theorem S* 2. Let $\mathcal{V} = \{v_1, \cdots, v_n\} \subset \mathbb{R}$ be $\gamma$−spherical and $\gamma$−well-separated clusterable data with $\mathcal{C}_1, \cdots, \mathcal{C}_k$ satisfying Definition 3 defining the $k$ individual clusters of size at least $0.1(n/k)$, where $k = O(1)$. Choose $\tau_i^2 = \frac{\gamma}{4} \min_{j\in[n]\setminus\{i\}}(v_i - v_j)^2 (\forall i \in [n])$, $h = 1$ and any $\alpha$ satisfying $O(\frac{\sqrt{n} \log n}{\alpha}) = o(1)$, $O(\frac{\alpha}{n}) = o(1)$.

Let $\mathcal{Y}$ be the output of t-SNE (algorithm 1) after $T = \Theta(\frac{n \log n}{\alpha})$ iterations on input $\mathcal{V}$ with the above parameters. Then, with probability at least 0.99 over the choice of the initialization, $\mathcal{Y}_{(T)}$ is a full visualization of $\mathcal{V}$.

*Remark* 2. Theorem S2 indicated that, in the presence of heterogeneity in the surrogate, t-SNE can project the subgroup structure encoded in the dissimilarity matrix constructed from pairwise differences in NIECC, into a two-dimensional space. As noted in Cai and Ma [2022], during the t-SNE embedding step, the resulting clusters are reliable in terms of their membership, whereas the relative positions between clusters are not, due to random initialization. Consequently, clusters that appear adjacent in the low-dimensional embedding do not necessarily correspond to neighboring clusters in the original data space.

The above theorem characterizes the behavior of t-SNE under an oracle setting in which the NIECC $v_1, \ldots, v_n$ are observed without error. When these latent effects satisfy standard clusterability and separation conditions, the resulting t-SNE embedding yields a valid visualization that recovers the underlying subgroup structure. This result provides a theoretical foundation for visualizing heterogeneity in indirect treatment effects when the population-level structure is well defined.

In practice, the NIECC are unknown and must be estimated from the observed data. As a result, t-SNE is applied to estimated quantities $\hat{v}_1, \ldots, \hat{v}_n$, introducing sampling variability into the visualization procedure. To justify the use of t-SNE in this setting, it is therefore necessary to assess the impact of estimation error on the induced similarity structure. In the subsequent analysis, we show that the oracle visualization guarantees extend to the estimated setting, provided that the estimation error is sufficiently controlled relative to the underlying cluster separation.

*Theorem S* 3. Let $v_1, \ldots, v_n \in \mathbb{R}$ denote the true NIECC, and suppose that the set $\mathcal{V} = \{v_1, \ldots, v_n\}$ is $\gamma$-spherical and $\gamma$-well-separated clusterable with a ground-truth partition $\mathcal{C}_1, \ldots, \mathcal{C}_k$, where $k = O(1)$ and $|\mathcal{C}_\ell| \geq 0.1(n/k)$ for all $\ell$. Let $P = (p_{ij})$ denote the pairwise similarity matrix constructed from $\mathcal{V}$ as in Algorithm 1.

Let $\hat{v}_1, \ldots, \hat{v}_n$ be estimators of $v_1, \ldots, v_n$ satisfying $\max_{1\leq i\leq n} |\hat{v}_i - v_i| \leq \delta_n$ with probability tending to one, where $\delta_n \to 0$. Let $\hat{P} = (\hat{p}_{ij})$ denote the corresponding similarity matrix constructed from $\hat{v}_1, \ldots, \hat{v}_n$ using the same bandwidth parameters.

Assume that $\delta_n = o(\gamma)$. Then, with probability at least 0.99 over the random initialization, the t-SNE embedding $\hat{Y}^{(T)}$ obtained after $T = \Theta(n \log n/\alpha)$ iterations applied to $\hat{P}$ yields a full visualization of the latent cluster structure $\mathcal{C}_1, \ldots, \mathcal{C}_k$.

*Remark* 3. Theorem S3 demonstrates that when the original NIECC exhibits a true class structure, applying the t-SNE algorithm to its estimated values preserves the class structure after projection into a two-dimensional space, thereby yielding a full visualization.

---

**Algorithm 1** t-SNE Embedding with Early Exaggeration

---

**Input** Dataset $V = \{v_1, \ldots, v_n\} \subset \mathbb{R}$; bandwidth parameters $\{\tau_i\}_{i=1}^{n}$; exaggeration factor $\alpha > 0$; step size $h > 0$; number of iterations $T$.

**1:** Compute pairwise similarities $\{p_{ij}\}_{i \neq j}$ according to (9).

**2:** Initialize $y_{i,(0)} \sim \mathrm{Unif}([-0.01, 0.01]^2)$ for all $i \in [n]$.

**for** $t = 0, 1, \ldots, T-1$ **do**

**3:** Compute normalization constant:

$$Z_{(t)} = \sum_{i \neq j} \left(1 + \left\| y_{i,(t)} - y_{j,(t)} \right\|^2\right)^{-1}.$$

**4:** Define low-dimensional similarities:

$$q_{ij,(t)} = \frac{\left(1 + \left\| y_{i,(t)} - y_{j,(t)} \right\|^2\right)^{-1}}{Z_{(t)}}, \quad i \neq j.$$

**5:** Update embeddings:

$$y_{i,(t+1)} = y_{i,(t)} + h \sum_{j \neq i} \left(\alpha p_{ij} - q_{ij,(t)}\right) q_{ij,(t)} Z_{(t)} \left(y_{i,(t)} - y_{j,(t)}\right), \quad \forall i \in [n].$$

**end**

**6:Output:** 2D embedding: $Y_{(T)} = \{y_{1,(T)}, \ldots, y_{n,(T)}\} \subset \mathbb{R}^2$

---

# 3 Web Appendix C:Identify patient profile

## 3.1 Custom metric

We make the following rules to select the final clustering result:

For each pre-defined $k$ clusters, we note them as $C_1, \cdots, C_k$.

Step 1: We use a decision tree to fit $k$ clusters from the K-means clustering. We denote each terminal node of the decision tree leaf.

Step 2: Two Cox models are fitted,

$$\lambda(t) = \lambda_0(t) \exp\{\beta_1 W\}, \tag{13}$$

$$\lambda(t) = \lambda_0(t)\exp\{\beta_2 W + \beta_3\text{leaf} + \beta_4\text{leaf} * W\}. \tag{14}$$

Let $l_1$ and $l_2$ denote the maximized partial log-likelihoods of models (13) and (14), respectively. The likelihood ratio test statistic,

$$-2(l_2 - l_1) \rightarrow \chi^2(df), \tag{15}$$

where $df$ = number of leaves −1. $p^Y_{\text{leaf}}$ is the p value of model (15).

Step 3:Two linear models are fitted,

$$M = \beta_5 W, \tag{16}$$

$$M = \beta_6 W + \beta_7\text{leaf} + \beta_8\text{leaf} * W. \tag{17}$$

Let $l_3$ and $l_4$ denote the maximized likelihoods of models (16) and (17), respectively. The likelihood ratio test statistic,

$$-2(l_2 - l_1) \rightarrow \chi^2(df), \tag{18}$$

where $df$ = number of leaves −1. $p^M_{\text{leaf}}$ is the p value of model (18)

Step 3: The selection metric is:

$$metric = I(p^Y_{\text{leaf}} < p^{Y^*}) \times I(p^M_{\text{leaf}} < p^{M^*}) \times p^M_{\text{leaf}}, \tag{19}$$

where $p^{Y^*}$ and $p^{M^*}$ serve as the thresholds for determining the presence of heterogeneity in the data and is selected through a calibration procedure as detailed in (12) in the main text, making it a data-driven parameter. We select the profile with the smallest metric as the final output.

## 3.2 Explained by decision tree

When fitting a decision tree to approximate the cluster assignments obtained from K-means, we introduced several constraints to mitigate overfitting:

1. **No repeated splitting variables along a path.** The same covariate is not allowed to be used more than once along a single branch of the tree, which prevents repeated partitioning based on the same variable.

2. **Maximum tree depth.** This controls the maximum number of splits along a path and therefore limits the number of distinct covariates that can define a subgroup.

3. **Minimum leaf size.** A minimum number of observations is required for each terminal node, ensuring that the identified subgroups are sufficiently large and practically meaningful. In practice, extremely small subgroups are often not reasonable in applied settings.

Web Figure 1 provides a detailed description of the three rules above.

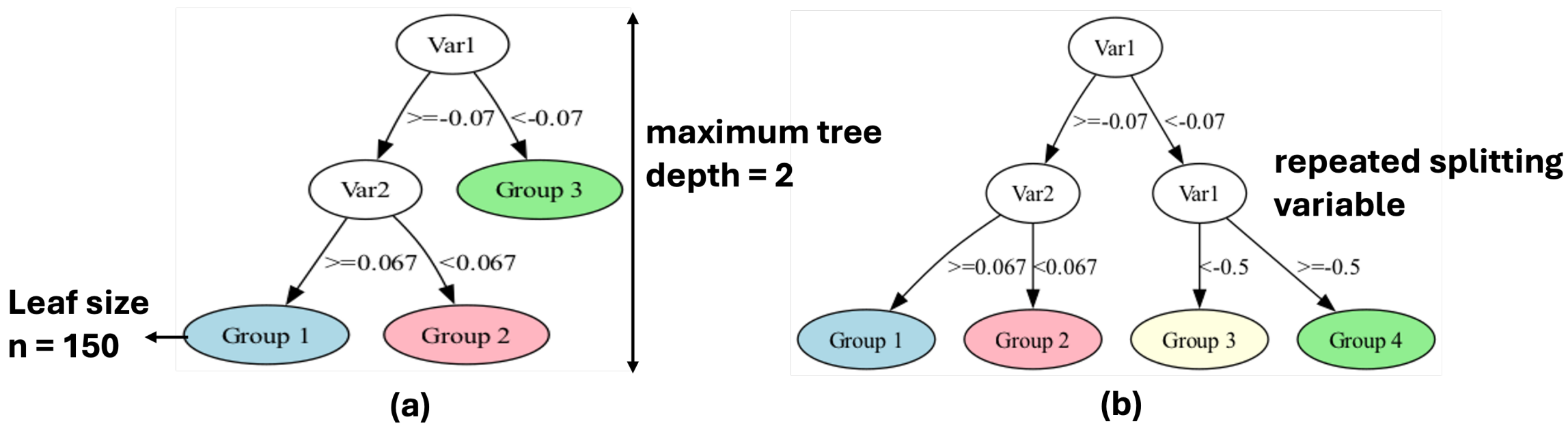

**Web Figure 1.** Illustration of the decision tree rules for explaining the identified clusters. (a) A decision tree satisfying the three rules: minimum leaf size of 150, maximum depth of 2, and no repeated splitting variables. (b) A decision tree that violates these rules; along two paths, var1 is used more than once.

In our simulation studies and real data analysis, we implemented constraints (1) and (3). The minimum leaf size was set to 200, meaning that each identified subgroup must contain at least 200 subjects.

## 3.3 Calibration

Because data-driven subgroup discovery methods actively search over a large space of candidate partitions, they were inherently prone to identifying spurious NIECC heterogeneity even when the true NIECC was homogeneous. In this setting, falsely declaring heterogeneity corresponded to a Type I error and represented a major concern for both statistical validity and clinical interpretability. To address this issue, we explicitly calibrated the heterogeneity detection procedure under the null scenario of no NIECC heterogeneity. Specifically, calibration was performed to ensure that the probability of detecting heterogeneity in homogeneous settings was properly controlled, thereby enabling a principled distinction between true NIECC heterogeneity and random fluctuations induced by data-driven model selection.

In our framework, NIECC heterogeneity was quantified using the statistic $p^{M}_{\text{leaf}}(p^{Y}_{\text{leaf}})$, with smaller values indicating stronger evidence against homogeneous NIECC. We declare the presence of heterogeneity when $p^{M}_{\text{leaf}} < p^{M^*}$ ($p^{Y}_{\text{leaf}} < p^{Y^*}$), where $p^{M^*}$ ($p^{Y^*}$) was a calibrated threshold. Owing to the data-driven construction of subgroups, the null distribution of $p^{M}_{\text{leaf}}(p^{Y}_{\text{leaf}})$ is not analytically tractable and must be obtained empirically.

To calibrate $p^{M^*}(p^{Y^*})$, we relied on simulation scenarios that are known to exhibit no NIECC heterogeneity. Specifically, both the null scenario and the global scenario corre-spond to homogeneous NIECC across patients, differing only in the magnitude of the overall treatment effect. Importantly, neither scenario contains subgroup-specific treatment effects; therefore, the corresponding $p_{\text{leaf}}$ values jointly characterize the behavior of the heterogeneity statistic under nonheterogeneous settings. We thus pooled the empirical distributions of $p_{\text{leaf}}$ obtained from the null and global scenarios to construct a robust empirical null distribution. The threshold $p^{M^*}(p^{Y^*})$ was selected as the 5st percentile of this pooled distribution (see (19)), such that, under true NIECC homogeneity, the probability of falsely declaring heterogeneity is controlled at 5%. This calibration ensures a Type I error rate of 5%, meaning that only 5% of truly non-heterogeneous cases would be incorrectly classified as exhibiting NIECC heterogeneity. By fixing the false positive rate at a pre-specified level, the proposed calibration yields a principled and reproducible decision rule for heterogeneity detection in data-driven subgroup analyses. Appendix Figure 3 and Figure 3 in the main text presented the corresponding calibration plots.

# 4 Web Appendix D: Proof of Theorem 1 in the main text

Under model for the mediator and model for the outcome:

$$\lambda_T(t|W=w, M=m, X=x) = \lambda_0(t)\exp\{\kappa_1(x)+\kappa_2(x)w+\kappa_3(x)m\}. \tag{20}$$

we have that

$$\lambda_{T_{w,M_{w^*}}}(t|X=x) = \frac{f_{T_{w,M_{w^*}}}(t|X=x)}{S_{T_{w,M_{w^*}}}(t|X=x)}, \tag{21}$$

where $f_{T_{w,M_{w^*}}}(t|X=x)$ and $S_{T_{w,M_{w^*}}}(t|X=x)$ denote the conditional density and survival functions respectively for $T_{w,M_{w^*}}$. We have that

$$\begin{aligned}
f_{T_{w,M_{w^*}}}(t|X=x) &= \int f_{T_{w,m}}(t|X=x, M_{w^*=m})dP_{M_{w^*}}(m|X=x)\\
&= \int f_{T_{w,m}}(t|X=x)dP_{M_{w^*}}(m|X=x)(\text{by Assumption 2})\\
&= \int f_T(t|X=x, M=m, W=w)dP_{M_{w^*}}(m|X=x)(\text{by Assumption 1, 2, 3})\\
&= \int \lambda_0(t)\exp\left(\kappa_1(x)+\kappa_2(x)w+\kappa_3(x)m\right)\\
&\qquad \exp\left(-\Lambda_T(t)\exp\left(\kappa_1(x)+\kappa_2(x)w+\kappa_3(x)m\right)\right)dP_{M_{w^*}}(m|X=x),
\end{aligned} \tag{22}$$

where $\Lambda_T(t) = \int_0^t \lambda_0(t)dt$. Likewise,

$$S_{T_{w,M_{w^*}}}(t|X=x) = \int \exp\left(-\Lambda_T(t)\exp\left(\kappa_1(x)+\kappa_2(x)w+\kappa_3(x)m\right)\right)dP_{M_{w^*}}(m|X=x). \tag{23}$$

Since,

$$
\begin{aligned}
& \lambda_{T_{w,M_{w^*}}}(t|X=x) \\
= \; & \frac{f_{T_{w,M_{w^*}}}(t|X=x)}{S_{T_{w,M_{w^*}}}(t|X=x)} \\
= \; & \frac{\int \lambda_0(t)\exp\left(\kappa_1(x)+\kappa_2(x)w+\kappa_3(x)m\right)\exp\left(-\Lambda_T(t)\exp\left(\kappa_1(x)+\kappa_2(x)w+\kappa_3(x)m\right)\right)dP_{M_{w^*}}(m|X=x}{\int \exp\left(-\Lambda_T(t)\exp\left(\kappa_1(x)+\kappa_2(x)w+\kappa_3(x)m\right)\right)dP_{M_{w^*}}(m|X=x)} \\
\approx \; & \int \lambda_0(t)\exp\left(\kappa_1(x)+\kappa_2(x)w+\kappa_3(x)m\right)dP_{M_{w^*}}(m|X=x),
\end{aligned} \tag{24}
$$

The last equation is an approximation by assuming the outcome is rare and then

$$
\exp\left(-\Lambda_T(t)\exp\left(\kappa_1(x)+\kappa_2(x)w+\kappa_3(x)m\right)\right) \approx 1.
$$

Hence,

$$
\begin{aligned}
\lambda_{T_{w,M_{w^*}}}(t|X=x) \;\approx\; & \int \lambda_0(t)\exp\left(\kappa_1(x)+\kappa_2(x)w+\kappa_3(x)m\right)dP_{M_{w^*}}(m|X=x) \\
= \; & \lambda_0(t)\exp(\{\kappa_1(x)+\kappa_2(x)w+\kappa_3(x))(\kappa_4(x)+\kappa_5(x)w^*) \\
& +\frac{1}{2}\kappa_3(x)^2\sigma_M^2) \quad .
\end{aligned}
$$

The last equation utilizes the property of moment generating function of normal random variable since $M_{w^*} \sim N(\kappa_3(x)(\kappa_4(x)+\kappa_5(x), \sigma_M^2)$ So,

$$
\log\lambda_{T_{w,M_{w^*}}} = \log(\lambda_0(t))+\kappa_1(x)+\kappa_2(x)w+\kappa_3(x)(\kappa_4(x)+\kappa_5(x)w^*)+\frac{1}{2}\kappa_3(x)^2\sigma_M^2 . \tag{25}
$$

From this it follows that,

$$
\begin{aligned}
\log\lambda_{T_{w,M_w}}(t|X=x)-\log\lambda_{T_{w,M_{w^*}}}(t|X=x) &= \kappa_3(x)\kappa_5(x)(w-w^*) \\
\log\lambda_{T_{w,M_{w^*}}}(t|X=x)-\log\lambda_{T_{w^*,M_{w^*}}}(t|X=x) &= \kappa_2(x)(w-w^*)
\end{aligned} \tag{26}
$$

Since we focus on the natural direct and indirect effects conditional on $W=1$, the treatment indicator in the outcome model is fixed at $w=1$, while $w^*$ only indexes the mediator distribution. Hence, $w=1$ and $w^*=0$ can be directly substituted in the derivations.

$$
\begin{aligned}
& \kappa_3(x)\kappa_5(x)(w-w^*) \\
= \; & (\kappa_3(x)\kappa_5(x)w+\kappa_3(x)\kappa_4(x)+\kappa_2(x)w+\kappa_1(x))-(\kappa_3(x)\kappa_5(x)w^*+\kappa_3(x)\kappa_4(x)+\kappa_2(x)w+\kappa_1(x)) \\
= \; & g(x|W=1, M=\mathrm{E}[M|X=x, W=w])-g(x|W=1, M=\mathrm{E}[M|X=x, W=w^*]),
\end{aligned} \tag{27}
$$

Since we focus on the natural direct and indirect effects conditional on $W=1$, the treatment variable in the outcome model is evaluated at $w=1$, while $w^*$ only indexes the intervention level of the mediator distribution. Consequently, the derivations can be carried out by setting $w=1$ and $w^* \in \{0, 1\}$.

# 5 Web Appendix E: Proof of Theorem S2

We break the proof of Theorem S1 into two parts: (i) Lemma S1, which identifies sufficient conditions on the pairwise affinities $p_{ij}$'s that imply that t-SNE outputs a full visualization, and (ii) Lemma S2, which shows that the $p_{ij}$'s computed for $\gamma$ -spherical, $\gamma$-well-separated data satisfy the requirements in Lemma S1. Combining Lemma S1 and S2 gives Theorem S2.

*Lemma S* 1. Consider a dataset $\mathrm{V} = \{v_1, v_2, \cdots, v_n\} \subseteq \mathrm{R}$ with ground truth clusters $\mathrm{C}_1, \cdots, \mathrm{C}_k$ satisfying $|C_l| \geq 0.1 * n/k$ for each $l \in [k]$ and $k = O(1)$. Let $\{p_{ij} : i, j \in [n], i \neq j\}$ be the pairwise affinities in t-SNE computed by (9). Suppose that there exist $\delta, \epsilon, \eta > 0$ such that $p_{ij}$, $\alpha$, and $h$ in t-SNE (algorithm 1) satisfy:

(1) for any cluster $l$, and any point $i \in \mathrm{C}_l$, we have $|\{j \in \mathrm{C}_l : \alpha h p_{ij} \geq \frac{\delta}{|\mathrm{C}_l|}\}| \geq (\frac{1}{2} + \eta)|\mathrm{C}_l|$;

(2) for any cluster $l$, and any point $i \in \mathrm{C}_l$, we have $\alpha h \sum_{j \in \mathrm{C}_l \setminus \{i\}} p_{ij} \leq 1$;

(3) for any cluster $l$, and any point $i \in \mathrm{C}_l$, we have $\alpha h \sum_{j \in \mathrm{C}_l \setminus \{i\}} p_{ij} + \frac{h}{n} \leq \epsilon$;

(4) $\frac{\sqrt{n}\epsilon \log 1/\epsilon}{\delta\eta} = o(1)$.

Then, with high probability over the choice of the random initialization, the output $\mathrm{Y}_{(T)}$ of t-SNE after $T = \Theta(\frac{\log\frac{1}{\epsilon}}{\delta\eta})$ iterations is a full visualization of $\mathrm{V}$.

*Lemma S* 2. Let $\mathrm{V} = \{v_1, v_2, \cdots, v_n\} \subset \mathrm{R}$ be $\gamma$ -spherical, $\gamma$-well-separated clusterable data with $\mathrm{C}_1, \mathrm{C}_2, \cdots, \mathrm{C}_k$ defining the individual clusters such that $|\mathrm{C}_i| \geq 0.1n/k$ for each $i$. Let $p_{i,j}$'s be the affinities computed by t-SNE (algorithm 1) with parameters $\tau_i^2 = \frac{\gamma}{4}\min_{j\in[n]\setminus\{i\}}(\tau_i - \tau_j)^2 (\forall i \in [n])$, $h = 1$, and any $\alpha$ satisfying $\alpha = o(n)$, and $\sqrt{n}\log n = o(\alpha)$.

Then, $p_{ij}$'s satisfy (1)-(4) in Lemma S1 with $\delta = \Theta(\alpha/n)$, $\epsilon = 2/n$ and $\eta = 0.01$.

## 5.1 Proof of Lemma S1

The proof of Lemma S1 is naturally divided into two parts. In the first part, we establish that over the course of the iterative updates in the t-SNE algorithm, distances between points in the same cluster decrease (Lemma S3). This step is essentially established in the work of [Linderman and Steinerberger, 2019], which we cite directly here. For completeness, we provide the proof in subsection 5.3.

*Lemma S* 3. Under the same setting as Lemma S1, after runing t-SNE for $T = \Theta(\frac{\log\frac{1}{\epsilon}}{\delta\eta})$ rounds, we have $\mathrm{Diam}(\{y_{i,(T)} : i \in \mathrm{C}_l\}) = O(\frac{\epsilon}{\delta\eta})$ for all $l \in [k]$.

In the second part, we establish that points in different clusters remain separated in the embedding if the clusters are well-separated in the input data. Concretely, let $\mu_{l,(t)} := \frac{1}{|\mathrm{C}_l|}\sum_{i\in\mathrm{C}_l} y_{i,(t)}$, which is the centroid of $\{y_{i,(t)} : i \in \mathrm{C}_l\}$. The following lemma says that the centroids of all clusters will remain separated in the first $O(\frac{\log\frac{1}{\epsilon}}{\delta\eta})$ rounds.

*Lemma S4.* Under the same setting as Lemma S1, if t-SNE is run for $T = O(\frac{\log\frac{1}{\epsilon}}{\delta\eta})$ iterations, with high probability we have $\|\mu_{l,(T)} - \mu_{l',(T)}\| = \Omega(\frac{1}{\sqrt{n}})$ for all $l \neq l'$.

We can finish the proof of Lemma S1 using the above two lemmas.

Using Lemmas S3 and S4, we know that after $T = \Theta(\frac{\log\frac{1}{\epsilon}}{\delta\eta})$ iterations, for any $i, j \in [n]$ we have:

if $i \sim j$, then $\|y_{i,(T)} - y_{j,(T)}\| \leq \mathrm{Diam}\,(\{y_{l,(t)} : l \in \mathcal{C}_{\pi(i)}\}) = O(\frac{\epsilon}{\delta\eta})$;

if $i \nsim j$, then $\|y_{i,(T)} - y_{j,(T)}\| \geq \|\mu_{\pi(i),(T)} - \mu_{\pi(j),(T)}\| - \|y_{i,(T)} - \mu_{\pi(i),(T)}\| - \|y_{j,(T)} - \mu_{\pi(j),(T)}\| \geq \|\mu_{\pi(i),(T)} - \mu_{\pi(j),(T)}\| - \mathrm{Diam}\{y_{l,(t)} : l \in \mathcal{C}_{\pi(i)}\} - \mathrm{Diam}\{y_{l,(t)} : l \in \mathcal{C}_{\pi(j)}\} \geq \Omega(\frac{1}{\sqrt{n}}) - O(\frac{\epsilon}{\delta\eta}) - O(\frac{\epsilon}{\delta\eta}) = \Omega(\frac{1}{\sqrt{n}})$, so $(\frac{\frac{\epsilon}{\delta\eta}}{\|y_{i,(T)} - y_{j,(T)}\|}) = o(1)$. Thus, in $\mathcal{Y}_{(T)}$, each point is much closer to points from its own cluster than to points from other clusters. So we prove Lemma S1.

## 5.2 Proof of Lemma S4

The idea is to track the centroids of the points coming from the true clusters and show that they remain separated at the end of the algorithm. Towards this, we first show random initialization ensures that the cluster centroids are initially well-separated with high probability.

*Lemma S5.* Suppose $|\mathcal{C}_l| \geq 0.1(n/k)$ for all $l \in [k]$. If $y_{i,(0)}$'s are generated i.i.d from the uniform distribution over $[-0.01, 0.01]^2$, then with probability at least 0.99 we have $\|\mu_{l,(0)} - \mu_{l',(0)}\| = \Omega(\frac{1}{\sqrt{n}})$ for all $l \neq l'$. For notational simplicity, the superscript "$_{(0)}$" is suppressed in the proof of this lemma. For a vector $y$, denoted by $y^{(1)}$ its first coordinate. Then it suffices to prove $\mu_l^{(1)} - \mu_{l'}^{(1)} = \Omega(\frac{1}{\sqrt{n}})$ for all $l \neq l'$.

Consider a fixed $l \in [k]$. Note the $y_i^{(1)}$'s are i.i.d with the uniform distribution over $[-0.01, 0.01]$, which clearly has zero mean and finite second and third absolute moments. Since $\mu_l^{(1)} = \frac{1}{|\mathcal{C}_l|} \sum_{i \in \mathcal{C}_l} y_i^{(1)}$, using the Berry-Esseen theorem [Berry, 1941, Esseen, 1942], we know that

$$|F(x) - \Phi(x)| \leq O(1/\sqrt{|\mathcal{C}_l|}),$$

where $F$ is the cumulative distribution function (CDF) of $\frac{\mu_l^{(1)}\sqrt{|\mathcal{C}_l|}}{\vartheta}$ ($\vartheta$ is the standard deviation of the uniform distribution over $[-0.01, 0.01]$), $\Phi(x)$ is the CDF of standard normal distribution.

It follow that for any fixed $a \in \mathbb{R}$ and $b > 0$, we have

$$
\begin{aligned}
\mathbb{P}\left[ |\mu_l^{(1)} - a| \leq \frac{b}{\sqrt{|C_l|}} \right] &= \mathbb{P}\left[ \frac{\mu_l^{(1)}\sqrt{|C_l|}}{\sigma} - \frac{a\sqrt{|C_l|}}{\sigma} \leq \frac{b}{\sigma} \right] \\
&= F\left( \frac{a\sqrt{|C_l|} + b}{\sigma} \right) - F\left( \frac{a\sqrt{|C_l|} - b}{\sigma} \right) \\
&\leq \Phi\left( \frac{a\sqrt{|C_l|} + b}{\sigma} \right) - \Phi\left( \frac{a\sqrt{|C_l|} - b}{\sigma} \right) + O\left( \frac{1}{\sqrt{|C_l|}} \right) \\
&= \int_{\frac{a\sqrt{|C_l|}-b}{\sigma}}^{\frac{a\sqrt{|C_l|}+b}{\sigma}} \frac{1}{\sqrt{2\pi}} e^{-\frac{x^2}{2}} \, dx + O(\sqrt{k/n}) \\
&\leq \int_{\frac{a\sqrt{|C_l|}-b}{\sigma}}^{\frac{a\sqrt{|C_l|}+b}{\sigma}} \frac{1}{\sqrt{2\pi}} \, dx + O(\sqrt{k/n}) \\
&= \frac{1}{\sqrt{2\pi}} \frac{2b}{\sigma} + O(\sqrt{k/n}).
\end{aligned}
$$

From $k = O(1)$, we have $\sqrt{k/n} = o(1)$. Therefore letting $b$ be a sufficiently small constant, we can ensure

$$
\mathbb{P}\left[ |\mu_l^{(1)} - a| \leq \frac{b}{\sqrt{|C_l|}} \right] \leq \frac{1}{\sqrt{2\pi}} \frac{2b}{\sigma} + O(\sqrt{k/n}) \leq 0.01 \tag{28}
$$

for any $a \in \mathbb{R}$. For any $l' \neq l$, because $\mu_l^{(1)}$ and $\mu_{l'}^{(1)}$ are independent, we can let $a = \mu_{l'}^{(1)}$ in the last equation, which tell us

$$
\mathbb{P}\left[ \left| \mu_l^{(1)} - \mu_{l'}^{(1)} \right| \leq \frac{b}{\sqrt{|C_\ell|}} \right] \leq 0.01.
$$

The above inequality holds for any $l, l' \in [k]$ $(l \neq l')$. Taking a union bound over all $l$ and $l'$, we know that with probability at least 0.99 we have $|\mu_l^{(1)} - \mu_{l'}^{(1)}| \geq \frac{b}{\sqrt{|C_l|}} = \Omega(\frac{1}{\sqrt{n}})$ for all $l, l' \in [k]$ $(l \neq l')$ simultaneously.

The following lemma says that the centroid of each cluster will move no more than $\epsilon$ in each of the first $\frac{0.01}{\epsilon}$ iterations.

*Lemma S6.* Under the same setting as Lemma S1, for all $t \leq \frac{0.01}{\epsilon}$ and all $l \in [k]$ we have $\left\| \mu_{l,(t+1)} - \mu_{l,(t)} \right\| \leq \epsilon$.

To prove the Lemma S6, we need the following Lemma.

*Lemma S7* (Claim A.4 in [Arora et al., 2018]). Under the setting as Lemma S1, for all $t \leq \frac{0.01}{\epsilon}$, we have $y_{i,(t)} \in [-0.02, 0.02]^2$ and $\epsilon_{i,(t)} \leq \epsilon$ for all $i \in [n]$.

*Lemma S8* (Claim A.5 in [Arora et al., 2018]). Under the same setting as Lemma S1, for all $t \leq \frac{0.01}{\epsilon}$, we have $y_{i,(t)} \in [-0.02, 0.02]^2$ and $\|\epsilon_{i,(t)}\| \leq \epsilon$ for all $i \in [n]$, as well as $\|y_{i,(t)} - y_{j,(t)}\| \leq 0.06$, $0.9 \leq q_{ij,(t)} Z_{(t)} \leq 1$ and $\frac{0.9}{n(n-1)} \leq q_{ij,(t)} \leq \frac{1}{0.9n(n-1)}$ for all $i, j \in [n]$ $(i \neq j)$.

Now, we start to prove the Lemma S6. Taking the average of (12) for all $i \in \mathbb{C}_l$, we obtain

$$
\begin{aligned}
\frac{1}{|\mathbb{C}_l|} \sum_{i \in \mathbb{C}_l} y_{i,(t+1)} &= \frac{1}{|\mathbb{C}_l|} \sum_{i \in \mathbb{C}_l} y_{i,(t)} + \frac{h}{|\mathbb{C}_l|} \sum_{i \in \mathbb{C}_l} \sum_{j \neq i} \left(\alpha p_{ij} - q_{ij,(t)}\right) q_{ij,(t)} Z_{(t)} \left(y_{j,(t)} - y_{i,(t)}\right) \\
&= \frac{1}{|\mathbb{C}_l|} \sum_{i \in \mathbb{C}_l} y_{i,(t)} + \frac{h}{|\mathbb{C}_l|} \sum_{i \in \mathbb{C}_l} \sum_{j \in \mathbb{C}_l, j \neq i} \left(\alpha p_{ij} - q_{ij,(t)}\right) q_{ij,(t)} Z_{(t)} \left(y_{j,(t)} - y_{i,(t)}\right) \\
&\quad + \frac{h}{|\mathbb{C}_l|} \sum_{i \in \mathbb{C}_l} \sum_{j \notin \mathbb{C}_l} \left(\alpha p_{ij} - q_{ij,(t)}\right) q_{ij,(t)} Z_{(t)} \left(y_{j,(t)} - y_{i,(t)}\right) \\
&= \frac{1}{|\mathbb{C}_l|} \sum_{i \in \mathbb{C}_l} y_{i,(t)} + \frac{h}{|\mathbb{C}_l|} \sum_{i,j \in \mathbb{C}_l, i \neq j} \left(\alpha p_{ij} - q_{ij,(t)}\right) q_{ij,(t)} Z_{(t)} \left(y_{j,(t)} - y_{i,(t)}\right) \\
&\quad + \frac{h}{|\mathbb{C}_l|} \sum_{i \in \mathbb{C}_l} \sum_{j \notin \mathbb{C}_l} \left(\alpha p_{ij} - q_{ij,(t)}\right) q_{ij,(t)} Z_{(t)} \left(y_{j,(t)} - y_{i,(t)}\right) \\
&= \frac{1}{|\mathbb{C}_l|} \sum_{i \in \mathbb{C}_l} y_{i,(t)} + \frac{h}{|\mathbb{C}_l|} \sum_{i \in \mathbb{C}_l} \sum_{j \notin \mathbb{C}_l} \left(\alpha p_{ij} - q_{ij,(t)}\right) q_{ij,(t)} Z_{(t)} \left(y_{j,(t)} - y_{i,(t)}\right).
\end{aligned}
$$

Thus we have

$$
\begin{aligned}
\left\| \mu_{l,(t+1)} - \mu_{l,(t)} \right\| &= \left\| \frac{h}{|\mathbb{C}_l|} \sum_{i \in \mathbb{C}_l} \sum_{j \notin \mathbb{C}_l} \left(\alpha p_{ij} - q_{ij,(t)}\right) q_{ij,(t)} Z_{(t)} \left(y_{j,(t)} - y_{i,(t)}\right) \right\| \\
&\leq \frac{h}{|\mathbb{C}_l|} \sum_{i \in \mathbb{C}_l} \sum_{j \notin \mathbb{C}_l} \left(\alpha p_{ij} + q_{ij,(t)}\right) q_{ij,(t)} Z_{(t)} \left\| y_{j,(t)} - y_{i,(t)} \right\|.
\end{aligned}
$$

Since $t \leq \frac{0.01}{\epsilon}$, we can apply Lemma S8 and get

$$
\begin{aligned}
\left\|\mu_{l,(t+1)} - \mu_{l,(t)}\right\| &= \left\| \frac{h}{|C_l|} \sum_{i\in C_l}\sum_{j\notin C_l} \left(\alpha p_{ij} - q_{ij,(t)}\right) q_{ij,(t)} Z_{(t)} \left(y_{j,(t)} - y_{i,(t)}\right) \right\| \\
&\leq \frac{h}{|C_l|} \sum_{i\in C_l}\sum_{j\notin C_l} \left(\alpha p_{ij} + q_{ij,(t)}\right) q_{ij,(t)} Z_{(t)} \left\|y_{j,(t)} - y_{i,(t)}\right\| \\
&\leq \frac{h}{|C_l|} \sum_{i\in C_l}\sum_{j\notin C_l} \left(\alpha p_{ij} + \frac{1}{0.9n(n-1)}\right) \cdot 1 \cdot 0.06 \\
&\leq \frac{1}{|C_l|} \sum_{i\in C_l} \left(\alpha h \sum_{j\notin C_l} p_{ij} + \frac{h}{0.9n}\right) \cdot 0.06 \\
&\leq \frac{0.06}{|C_l|} \sum_{i\in C_l} \frac{\epsilon}{0.9} \\
&\leq \epsilon
\end{aligned}
$$

where we have used $\alpha h \sum_{j\notin C_l} p_{ij} + \frac{h}{n} \leq \epsilon$ for all $i \in C_l$ (condition 3 in Lemma S1).

Using Lemmas S5 and S6, we can complete the proof of Lemma S4.

Notice that we have $T = O(\frac{\log\frac{1}{\epsilon}}{\delta\eta}) \ll \frac{1}{\sqrt{n\epsilon}} < \frac{0.01}{\epsilon}$, where we have used condition (4) in Lemma S1. Hence we can apply Lemma S6 for all $t \leq T$. Lemma S6 says that after each iteration every centroid moves by at most $\epsilon$. This means that the distance between any two centroids changes by at most $2\epsilon$. Since with high probability the initial distance between $\mu_{l,(0)}$ and $\mu_{l',(0)}$ is at least $\Omega(\frac{1}{\sqrt{n}})$ (Lemma S5), we know that after $T$ rounds we have $\left\|\mu_{l,(T)} - \mu_{l',(T)}\right\| \geq \Omega(\frac{1}{\sqrt{n}}) - 2T\epsilon = \Omega(\frac{1}{\sqrt{n}}) - O(\frac{\epsilon\log\frac{1}{\epsilon}}{\delta\eta})$ with high probability, where the last step is due to condition 4 in Lemma S1.

## 5.3 Proof of Lemma S2

For each $i \in [n]$, define $A_i := \{j \sim i, j \neq i : (v_i - v_j)^2 \leq b_{\pi(i)}\}$, where $b_{\pi(i)}$ is the same as in Definition 3. We know $|A_i| \geq 0.51|C_{\pi(i)}|$.

We first to show that there exist constants $c_1, c_2 > 0$ such that for any $i, j \in [n] (i \neq j)$, we have

$$
p_{ij} \begin{cases} \leq \frac{c_2}{|C_{\pi(i)}|\, n}, & i \sim j, \\ \in \left[\frac{c_1}{|C_{\pi(i)}|\, n}, \frac{c_2}{|C_{\pi(i)}|\, n}\right], & j \in A_i, \\ \leq \frac{1}{n^3}, & i \not\sim j. \end{cases} \tag{29}
$$

Define $a_l := \frac{b_l}{1+\gamma}$ for all $l \in [k]$. Consider any fixed $i \in [n]$. Suppose $i \in C_l$. From the $\gamma$–sphericalness and $\gamma$–well-separation properties we know $a_l \leq \min_{j\in[n]\setminus\{i\}}(v_i - v_j)^2 \leq b_l$, and thus $\frac{\gamma}{4}a_l \le v_i^2 \le \frac{\gamma}{4}b_l$. Let $N_i = \sum_{j\in[n]\setminus\{i\}} \exp(-(v_i - v_j)^2/2v_i^2)$. Recall that from (9) we have $p_{j|i} = \frac{\exp(-(v_i-v_j)^2/2v_i^2)}{N_i}$ for all $j \in [n] \setminus \{i\}$. Then we have

$$
\begin{aligned}
p_{j|i} &\le \frac{\exp(-a_l/2\tau_i^2)}{N_i}, \quad \forall j \sim i,\\
\frac{\exp(-b_l/2\tau_i^2)}{N_i} \le p_{j|i} &\le \frac{\exp(-a_l/2\tau_i^2)}{N_i}, \quad \forall j \in \mathrm{A}_i,\\
p_{j|i} &\le \frac{\exp(-(1+\gamma\log n)b_l/2\tau_i^2)}{N_i}, \quad \forall j \not\sim i.
\end{aligned} \tag{30}
$$

Note that we have

$$
\frac{\exp(-a_l/2v_i^2)}{\exp(-b_l/2v_i^2)} = \exp\left(\frac{b_l - a_l}{2v_i^2}\right) = \exp\left(\frac{\gamma a_l}{2v_i^2}\right) \le \exp\left(\frac{\gamma a_l}{\frac{\gamma}{2}a_l}\right) = e^2
$$

and

$$
\frac{\exp(-b_l/2v_i^2)}{\exp(-(1+\gamma\log n)b_l/2v_i^2)} = \exp\left(\frac{\gamma\log n \cdot b_l}{2v_i^2}\right) \ge \exp\left(\frac{\gamma\log n \cdot b_l}{\frac{\gamma}{2}b_l}\right) = n^2
$$

As a consequence, letting $f_i := \frac{\exp(-b_l/2v_i^2)}{N_i}$, we know that (30) implies

$$
p_{j|i} \le O(f_i), \quad \forall j \sim i, \tag{31}
$$

$$
f_i \le p_{j|i} \le O(f_i), \quad \forall j \in \mathrm{A}_i, \tag{32}
$$

$$
p_{j|i} \le \frac{f_i}{n^2}, \quad \forall j \not\sim i. \tag{33}
$$

Since $\sum_{j\in[n]\setminus\{i\}} p_{j|i} = 1$, from (31) we have

$$
1 = \sum_{j\in[n]\setminus\{i\}} p_{j|i} = \sum_{j\sim i, j\neq i} p_{j|i} + \sum_{j\not\sim i} p_{j|i} \le \sum_{j\sim i, j\neq i} O(f_i) + \sum_{j\not\sim i} \frac{f_i}{n^2} \le O(|\mathrm{C}_l| f_i) + \frac{f_i}{n} = O(|\mathrm{C}_l| f_i)
$$

and

$$
1 = \sum_{j\in[n]\setminus\{i\}} p_{j|i} \ge \sum_{j\in\mathrm{A}_i} p_{j|i} \ge |\mathrm{A}_i| f_i \ge \frac{1}{2}|\mathrm{C}_l| f_i
$$

Thus we have $f_i = O(\frac{1}{|C_l|})$. Then (31) leads to

$$
p_{j|i} \begin{cases} = O\left(\frac{1}{|C_l|}\right), & \forall j \sim i,\\ = \Theta\left(\frac{1}{|C_l|}\right), & \forall j \in \mathrm{A}_i,\\ \le \frac{1}{n^2}, & \forall j \not\sim i. \end{cases} \tag{34}
$$

plugging this into (9), we obtain the desired bounds on $p_{ij}$'s:

$$p_{j|i} \begin{cases} = O\left(\frac{1}{|C_l|n}\right), & \forall j \sim i, \\ = \Theta\left(\frac{1}{|C_l|n}\right), & \forall j \in A_i, \\ \leq \frac{1}{n^3}, & \forall j \not\sim i. \end{cases} \tag{35}$$

Therefore, (29) is proved. Next, based on (29), we verify that the four conditions in Lemma S1 are all satisfied with parameters $\delta = \frac{c_1\alpha}{n}$, $\epsilon = \frac{2}{n}$ and $\eta = 0.01$.

Consider any $i \in [n]$, for all $j \in A_i$, we have $\alpha h p_{ij} \geq \frac{\alpha c_1}{C_{\pi(i)}n} = \frac{\delta}{\|C_{\pi(i)}\|}$. Since $A_i \geq (\frac{1}{2} + \eta)\|C_{\pi(i)}\|$, we have verified condition 2 in Lemma S1.

We have $\alpha h \sum_{j\sim i, j\neq i} p_{ij} \leq \alpha|C_{\pi(i)}| \cdot \frac{c_2}{|C_{\pi(i)}|n} = \frac{c_2\alpha}{n} \leq 1$, where we have used $O(\frac{\alpha}{n}) = o(1)$. Hence condition 2 is verified. We have $\alpha h \sum_{j\sim i, j\neq i} p_{ij}p_{ij} + \frac{h}{n} \leq \alpha n \frac{1}{n^3} + \frac{1}{n} \leq \frac{2}{n} = \epsilon$, which verifies condition 3. Finally, we have $\frac{\epsilon \log \frac{1}{\epsilon}}{\delta\eta} = O(\frac{\log n}{\alpha})$, $\frac{k^2\sqrt{n}\log n}{\alpha} = o(1)$, which verifies condition 4. Therefore we have finished the proof of Lemma S2.

## 5.4 Proof of Lemma S3

*Lemma S*9 (Lemma 1 in [Linderman and Steinerberger, 2019]). Let $z_1, \cdots, z_m \in \mathbb{R}^s$ be evolving as the following dynamic system:

$$z_{i,(t+1)} = \sum_{j=1}^{m} \lambda_{ij,(t)} z_{j,(t)} + \epsilon_{i,(t)}, \; i \in [m], \; t = 0, 1, 2, \cdots \tag{36}$$

and let $z_{i,(t)}$ := the position of $z_i$ at time $t$. Denote by $Conv_{(t)}$ the convex hull of $z_{1,(t)}, \cdots, z_{m,(t)}$, where $D_{(t)}$ is the Diam($Conv_{(t)}$). Suppose at time $t$ we have $\lambda_{ij,(t)} \geq 0 (\forall i, j \in [m])$, $\sum_{j=1}^{m} \lambda_{ij,(t)} = 1(\forall i \in [m])$; $\|\epsilon_{i,(t)}\| \leq \epsilon \; (\forall i \in [m])$. Then we have $Conv_{(t+1)} \subseteq Conv_{(t)} + \mathbb{B}(0, \epsilon)$.

*Lemma S*10 (Lemma A.2 in [Arora et al., 2018]). Under the same setting as Lemma S9, if furthermore there exist $\delta', \eta' > 0$ such that:

$$|\{j : \lambda_{ij,(t)} \geq \delta'\}| \geq (\frac{1}{2} + \eta')m, \forall i \in [m],$$

then we have $D_{(t+1)} \leq (1 - m\delta'\eta'/2)D_{(t)} + 2\epsilon'$

*Lemma S*11 (Corollary A.3 in [Arora et al., 2018]). Under the same setting as Lemma S10, we have:(1) if $D_{(t)} \geq \frac{5\epsilon}{m\delta'\eta'}$, then $D_{(t+1)} \leq (1 - m\delta'\eta'/2)D_{(t)}$;(2)if $D_{(t)} \leq \frac{5\epsilon}{m\delta'\eta'}$, then $D_{(t+1)} \leq \frac{5\epsilon}{m\delta'\eta'}$.

We rewrite the evolution of points in $\{y_{i,(t)} : i \in C_l\}$(for each $l \in [k]$) in the form of the

dynamic system red(11):

$$
\begin{aligned}
y_{i,(t+1)} &= y_{i,(t)} + \alpha h \sum_{j\sim i, j\neq i} p_{ij} q_{ij,(t)} Z_{(t)} \left( y_{j,(t)} - y_{i,(t)} \right) \\
&\quad + \alpha h \sum_{j\not\sim i} p_{ij} q_{ij,(t)} Z_{(t)} \left( y_{j,(t)} - y_{i,(t)} \right) - h \sum_{j\neq i} \left( q_{ij,(t)} \right)^2 Z_{(t)} \left( y_{j,(t)} - y_{i,(t)} \right) \\
&= \sum_{j\in \mathrm{C}_l} \lambda_{ij,(t)} y_{j,(t)} + \epsilon_{i,(t)}, \quad i \in \mathrm{C}_l,
\end{aligned} \tag{37}
$$

where

$$
\begin{aligned}
\lambda_{ij,(t)} &:= \alpha h p_{ij} q_{ij,(t)} Z_{(t)}, \quad i \sim j, i \neq j, \\
\lambda_{ii,(t)} &:= 1 - \sum_{j\sim i, j\neq i} \lambda_{ij,(t)}, \quad i \in [n], \\
\epsilon_{i,(t)} &:= \alpha h \sum_{j\not\sim i} p_{ij} q_{ij,(t)} Z_{(t)} \left( y_{j,(t)} - y_{i,(t)} \right) - h \sum_{j\neq i} \left( q_{ij,(t)} \right)^2 Z_{(t)} \left( y_{j,(t)} - y_{i,(t)} \right), \quad i \in [n].
\end{aligned}
$$

We can verify $\lambda_{ij,(t)} \geq 0$ for all $i \sim j, i \neq j$, and

$$
\lambda_{ii,(t)} = 1 - \sum_{j\sim i, j\neq i} \alpha h p_{ij} q_{ij,(t)} Z_{(t)} = 1 - \sum_{j\sim i, j\neq i} \alpha h p_{ij} \left( 1 + \left\| y_{i,(t)} - y_{j,(t)} \right\|^2 \right)^{-1} \geq 1 - \sum_{j\sim i, j\neq i} \alpha h p_{ij} \geq 0, \forall i \in [n],
$$

where the last inequality is due to Condition (3) in Lemma S1.

Now, we return to the proof of Lemma S3. We consider any time $t \leq \frac{0.01}{\epsilon}$. Consider any cluster $\mathrm{C}_l (l \in [k])$ and any $i, j \in \mathrm{C}_l (i \neq j)$. Using Lemma S8, we have:

$$
\alpha h p_{ij} \geq \frac{\delta}{|\mathrm{C}_l|} \Rightarrow \lambda_{ij,(t)} = \alpha h p_{ij} q_{ij,(t)} Z_{(t)} \geq \frac{0.9\delta}{|\mathrm{C}_l|}.
$$

Then from condition 1 in Lemma S1, we know that $|\{j \in \mathrm{C}_l : \lambda_{ij,(t)} \geq \frac{0.9\delta}{|\mathrm{C}_l|}\}| \geq (\frac{1}{2} + \eta)|\mathrm{C}_l|$ for all $i \in \mathrm{C}_l$. Recall that we also have $\left\| \epsilon_{i,(t)} \right\| \leq \epsilon (\forall i \in [n])$ according to Lemma S11. Hence the dynamic system (36) satisfy all the conditions in Lemma S10 and S11. using Lemma S11, we know $Diam(\{y_{i,(T)} : i \in \mathrm{C}_l\}) = O(\frac{\epsilon}{|\mathrm{C}_l| \frac{0.9\delta}{|\mathrm{C}_l|} \eta}) = O(\frac{\epsilon}{\delta\eta})$ as long as $T = \Omega(\frac{\log \frac{\delta\eta}{\epsilon}}{\delta\eta})$. This can be satisfied if $T = O(\frac{\log \frac{1}{\epsilon}}{\delta\eta})$ since $\delta\eta = O(1)$. Notice that we still need to check $T \leq \frac{0.01}{\epsilon}$ since otherwise we cannot use Lemma S7 this can be checked by observing $T = O(\frac{\log \frac{1}{\epsilon}}{\delta\eta}) = o(\frac{0.01}{\epsilon})$, where we have used Condition (4) in Lemma S1. Therefore, we have proved Lemma S3.

# 6 Web Appendix F: Proof of Theorem S3

We show that controlling the estimation error $\max_i |\hat{v}_i - v_i| \leq \delta_n$ implies uniform control of the induced t-SNE similarities and hence preserves the visualization guarantees.

Assume $|v_i| \leq B$ for all $i$ and $\tau_i \geq \tau_{\min} > 0$. Write $\hat{v}_i = v_i + e_i$ with $|e_i| \leq \delta_n$. For any $i \neq j$,

$$(\hat{v}_i - \hat{v}_j)^2 - (v_i - v_j)^2 = 2(v_i - v_j)(e_i - e_j) + (e_i - e_j)^2.$$

Taking absolute values and using $|v_i - v_j| \leq 2B$ yields

$$(\hat{v}_i - \hat{v}_j)^2 - (v_i - v_j)^2 \leq 4B\delta_n + 4\delta_n^2 \leq C_1\delta_n,$$

for some constant $C_1 > 0$.

Define $a_{ij} = \exp\left(-\frac{(v_i-v_j)^2}{2\tau_i^2}\right)$, $\hat{a}_{ij} = \exp\left(-\frac{(\hat{v}_i-\hat{v}_j)^2}{2\tau_i^2}\right)$. The function $f(x) = \exp(-x/(2\tau_i^2))$ is Lipschitz on bounded sets with $|f'(x)| \leq (2\tau_{\min}^2)^{-1}$. By the mean value theorem,

$$|\hat{a}_{ij} - a_{ij}| \leq \frac{C_1}{2\tau_{\min}^2}\delta_n =: C_2\delta_n.$$

Let $A_i = \sum_{\ell \neq i} a_{i\ell}$, $\hat{A}_i = \sum_{\ell \neq i} \hat{a}_{i\ell}$. Then $|\hat{A}_i - A_i| \leq \sum_{\ell \neq i} |\hat{a}_{i\ell} - a_{i\ell}| \leq (n-1)C_2\delta_n$. Moreover, since $|v_i| \leq B$ and $\tau_i \geq \tau_{\min}$, $a_{i\ell} \geq \exp\left(-\frac{(2B)^2}{2\tau_{\min}^2}\right) =: c_0 > 0$, and hence $A_i \geq (n-1)c_0$. Therefore, $\frac{|\hat{A}_i - A_i|}{A_i} \leq \frac{C_2}{c_0}\delta_n$.

The conditional probabilities are $p_{j|i} = a_{ij}/A_i$ and $\hat{p}_{j|i} = \hat{a}_{ij}/\hat{A}_i$. We decompose $\hat{p}_{j|i} - p_{j|i} = \frac{\hat{a}_{ij} - a_{ij}}{A_i} + \hat{a}_{ij}\left(\frac{1}{\hat{A}_i} - \frac{1}{A_i}\right)$. Using the bounds above and $\hat{a}_{ij} \leq 1$, we obtain $|\hat{p}_{j|i} - p_{j|i}| \leq C_3\delta_n$, for some constant $C_3 > 0$ independent of $i, j$.

Since

$$p_{ij} = \frac{p_{j|i} + p_{i|j}}{2n}, \qquad \hat{p}_{ij} = \frac{\hat{p}_{j|i} + \hat{p}_{i|j}}{2n},$$

it follows directly that

$$|\hat{p}_{ij} - p_{ij}| \leq C_4\delta_n.$$

The clusterability and separation conditions in Definition 3 depend only on the relative magnitudes of the entries of $P = (p_{ij})$. If $\delta_n = o(\gamma)$, then the uniform bound $\max_{i \neq j} |\hat{p}_{ij} - p_{ij}| \leq C_4\delta_n$ implies that $\hat{P}$ satisfies the same conditions with high probability, possibly with modified constants. Applying the t-SNE visualization theorem to $\hat{P}$ completes the proof.

# 7 Web Appendix G: Additional simulation details and results

## 7.1 Simulation details for linear assumption

Simulation studies were conducted with sample sizes $n = 1000$, reflecting realistic scenarios commonly encountered in clinical trial analyses. The total trial duration (maximum follow-up time) was set to 40 months for all scenarios. Patient enrollment was assumed to occur uniformly over the accrual period. Time-to-event outcomes were generated accordingly, with administrative censoring applied at the end of the trial. In addition, independent random censoring was incorporated and assumed to be independent of treatment assignment and potential outcomes.

For scenario 1 (Heterogeneous), the coefficients of $\beta_1, \beta_2, \beta_3, \beta_4$ were $-0.1, -0.1, 0.1, 0.1$ seperately, $\beta_5, \cdots, \beta_{10}, \beta_w$ were 0. The coefficient of $\beta_m$ was $-0.75$. In surrogate model, $\kappa_1(\boldsymbol{X}) = 0.1X^{(3)} + 0.1X^{(4)}$, and $\kappa_2(\boldsymbol{X}) = I(X^{(1)} > 0 \& X^{(2)} > 0)$, $\epsilon \sim N(0, 0.01)$. This scenario evaluates the proposed method under a setting where treatment effect heterogeneity is fully mediated through the surrogate and is aligned with the same covariate-defined subgroup, with a constant treatment effect within that subgroup.

For scenario 2 (Null), the coefficients of $\beta_1, \cdots, \beta_{10}, \beta_w$ were same to the Heterogeneous scenario. The coefficient of $\beta_m$ was 0. In surrogate model, $\kappa_1(\boldsymbol{X}) = 0.1X^{(3)} + 0.1X^{(4)}$, $\kappa_2(\boldsymbol{X}) = I(X^{(1)} > 0 \& X^{(2)} > 0)$, $\epsilon \sim N(0, 0.01)$, This scenario examines the ability of the method to avoid spurious findings when the surrogate exhibits treatment-related heterogeneity but does not mediate any treatment effect on the clinical outcome.

For scenario 3 (Global), the coefficients of $\beta_1, \cdots, \beta_{10}$ were same to the Heterogeneous scenario. The coefficient of $\beta_m$ was $-0.7$, and $\beta_w$ was 0. In surrogate model, $\kappa_1(\boldsymbol{X}) = 0.1X^{(3)} + 0.1X^{(4)}$, and $\kappa_2(\boldsymbol{X}) = 1$. $\epsilon \sim N(0, 0.01)$. This scenario assesses performance when the treatment effect is homogeneous across individuals and the surrogate reliably reflects the treatment effect on the outcome.

The mediator was estimated using two approaches. Under the KRLS method, estimation was conducted using the KRLS package (version 1.1.0) in R with default settings. Under the random forest method, estimation was performed using the randomForestSRC package (version 3.2.4) in R with default settings.

## 7.2 Simulation details for complex assumption

For scenario 1 (All1), $\kappa_1(\boldsymbol{X}) = -0.1X^{(1)} - 0.1X^{(2)} + 0.1X^{(3)} + 0.1X^{(4)}$, $\kappa_2(\boldsymbol{X}) = 0$, $\kappa_3(\boldsymbol{X}) = -0.75$, $\kappa_4(\boldsymbol{X}) = 0.1X^{(3)} + 0.1X^{(4)}$, $\kappa_5(\boldsymbol{X}) = I(X^{(1)} > 0 \& X^{(2)} > 0)$, $\epsilon \sim N(0, 0.01)$.

For scenario 2 (Part1), $\kappa_1(\boldsymbol{X}) = -0.1X^{(1)} - 0.1X^{(2)} + 0.1X^{(3)} + 0.1X^{(4)}$, $\kappa_2(\boldsymbol{X}) = -0.33I(X^{(1)} > 0 \& X^{(2)} > 0)$, $\kappa_3(\boldsymbol{X}) = -0.33$, $\kappa_4(\boldsymbol{X}) = 0.1X^{(3)} + 0.1X^{(4)}$, $\kappa_5(\boldsymbol{X}) = I(X^{(1)} > 0 \& X^{(2)} > 0)$, $\epsilon \sim N(0, 0.01)$.

For scenario 3 (Null), $\kappa_1(\boldsymbol{X}) = -0.1X^{(1)} - 0.1X^{(2)} + 0.1X^{(3)} + 0.1X^{(4)}$, $\kappa_2(\boldsymbol{X}) = 0$, $\kappa_3(\boldsymbol{X}) = 0$, $\kappa_4(\boldsymbol{X}) = 0.1X^{(3)} + 0.1X^{(4)} + I(X^{(1)} > 0 \& X^{(2)} > 0)$, $\kappa_5(\boldsymbol{X}) = 0$, $\epsilon \sim N(0, 0.01)$.

For scenario 4 (Global), $\kappa_1(\boldsymbol{X}) = -0.1X^{(1)} - 0.1X^{(2)} + 0.1X^{(3)} + 0.1X^{(4)}$, $\kappa_2(\boldsymbol{X}) = 0$, $\kappa_3(\boldsymbol{X}) = -0.75$, $\kappa_4(\boldsymbol{X}) = 0.1X^{(3)} + 0.1X^{(4)}$, $\kappa_5(\boldsymbol{X}) = (X^{(1)} > 0 \& X^{(2)} > 0)$, $\epsilon \sim N(0, 0.01)$.

For scenario 5 (All2), $\kappa_1(\boldsymbol{X}) = -0.1X^{(1)} - 0.1X^{(2)} + 0.1X^{(3)} + 0.1X^{(4)}$, $\kappa_2(\boldsymbol{X}) = 0$, $\kappa_3(\boldsymbol{X}) = -0.75$, $\kappa_4(\boldsymbol{X}) = \epsilon$, $\kappa_5(\boldsymbol{X}) = (X^{(1)} + X^{(2)})I(X^{(1)} > 0 \& X^{(2)} > 0)$, $\epsilon \sim N(0, 0.01)$.

For scenario 6 (Part2), $\kappa_1(\boldsymbol{X}) = -0.1X^{(1)} - 0.1X^{(2)} + 0.1X^{(3)} + 0.1X^{(4)}$, $\kappa_2(\boldsymbol{X}) = -0.2I(X^{(1)} > 0 \& X^{(2)} > 0)$, $\kappa_3(\boldsymbol{X}) = -0.2$, $\kappa_4(\boldsymbol{X}) = \epsilon$, $\kappa_5(\boldsymbol{X}) = (X^{(1)} + X^{(2)})I(X^{(1)} > 0 \& X^{(2)} > 0)$, $\epsilon \sim N(0, 0.01)$.

The mediator and log-hazard functions were estimated using the xgboost package (version 1.7.9.1) in R. For the estimation of the mediator, the tuning parameters were set to $\eta$ = 0.08, max_depth = 3, nrounds = 80, and subsample = 0.67. For the estimation of the log-hazard function, the parameters were set to $\eta$ = 0.06, max_depth = 2, nrounds = 80, and subsample = 0.67.

## 7.3 Simulation results

Web Figure 2 and Figure 5 present scatter plots obtained under the linear modeling assumption using KRLS and random forest methods, respectively. Simulation results indicate that, under the linear modeling assumption, the more flexible random forest approach for modeling the mediator is better suited to capturing heterogeneous effects than KRLS, as the subgroup boundary derived from the profile is visibly closer to the true boundary at $(0, 0)$.

Web Figure 3 and Figure 6 present the ecdf curves under $p^{M}_{leaf}$ and $p^{Y}_{leaf}$, using KRLS and random forest methods, respectively.

Web Figures Figure 4 and Figure 7 summarize the number of rejections based on the custom metric, the number of runs that produced a final output, and the frequency with which the variables Continuous 1 and Continuous 2 were selected, using KRLS and random forest, respectively. In the heterogeneous setting, selecting these variables corresponds to the correct result, whereas in the global and null settings, the absence of output represents the correct result.

Web Figure 8 presents the number of times the proposed method correctly selects the appropriate profile and rejects profiles under the complex assumption across different scenarios. In the heterogeneous setting, selecting these variables corresponds to the correct outcome, whereas in the global and null settings, the absence of any selected output represents the correct result.

Web Table 1 summarizes the bias and RMSE of NIECC estimates under different heterogeneous simulation scenarios. The results indicate that, compared with the complex modeling assumption, the linear assumption yields smaller bias and RMSE. However, the subgroup identification based on similarity exhibits larger estimation errors under the linear assumption than under the complex assumption.

These findings suggest that accurate estimation of bias and RMSE does not necessarily imply accurate subgroup identification. This discrepancy is further illustrated by the subgroup threshold plots, which reveal differences in the accuracy of the identified heterogeneous regions (Figure 2 and Figure 2 in the main text.).

**Web Table 1.** Summary of Bias and RMSE under Different Simulation Scenarios. Scenarios All1 and All2 correspond to complete mediation of the treatment effect through the mediator, whereas Part1 and Part2 correspond to partial mediation. The treatment effect is constant in All1 and Part1, and covariate-dependent in All2 and Part2. RF:random forest.

| Assumption | Scenario | bias | rMSE |
|---|---|---|---|
| Linear (RKLS) | Heterogeneous | 0.00 | 0.01 |
| Linear (RF) | Heterogeneous | 0.00 | 0.00 |
| Complex | All1 | 0.04 | 0.20 |
| Complex | Part1 | 0.06 | 0.16 |
| Complex | All2 | 0.09 | 0.28 |
| Complex | Part2 | 0.00 | 0.15 |

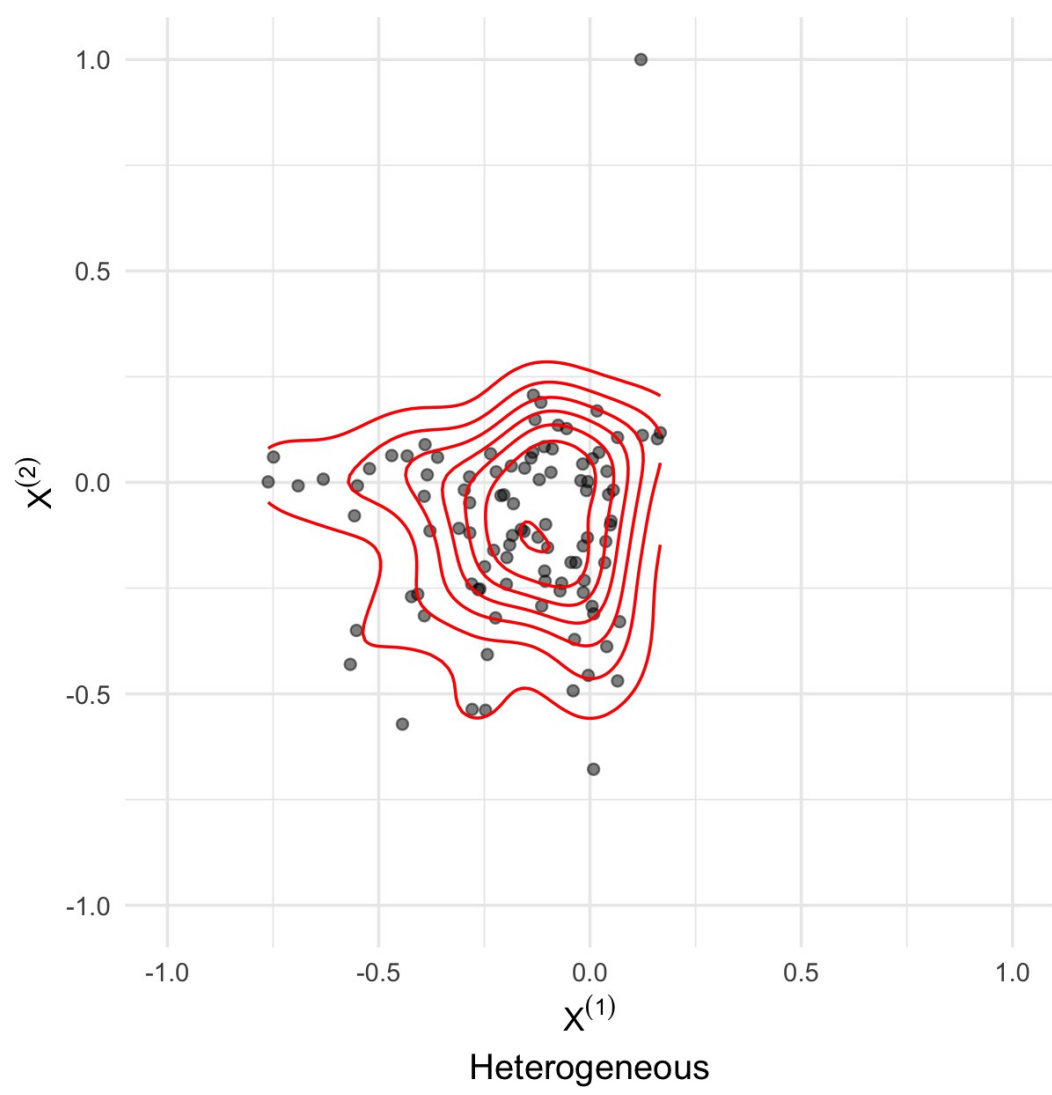

**Web Figure 2.** Distribution of thresholds of linear assumption in Heterogeneous scenario with RKLS method. Black points denote replicate-specific estimates, and the red curves represent the corresponding density estimates.

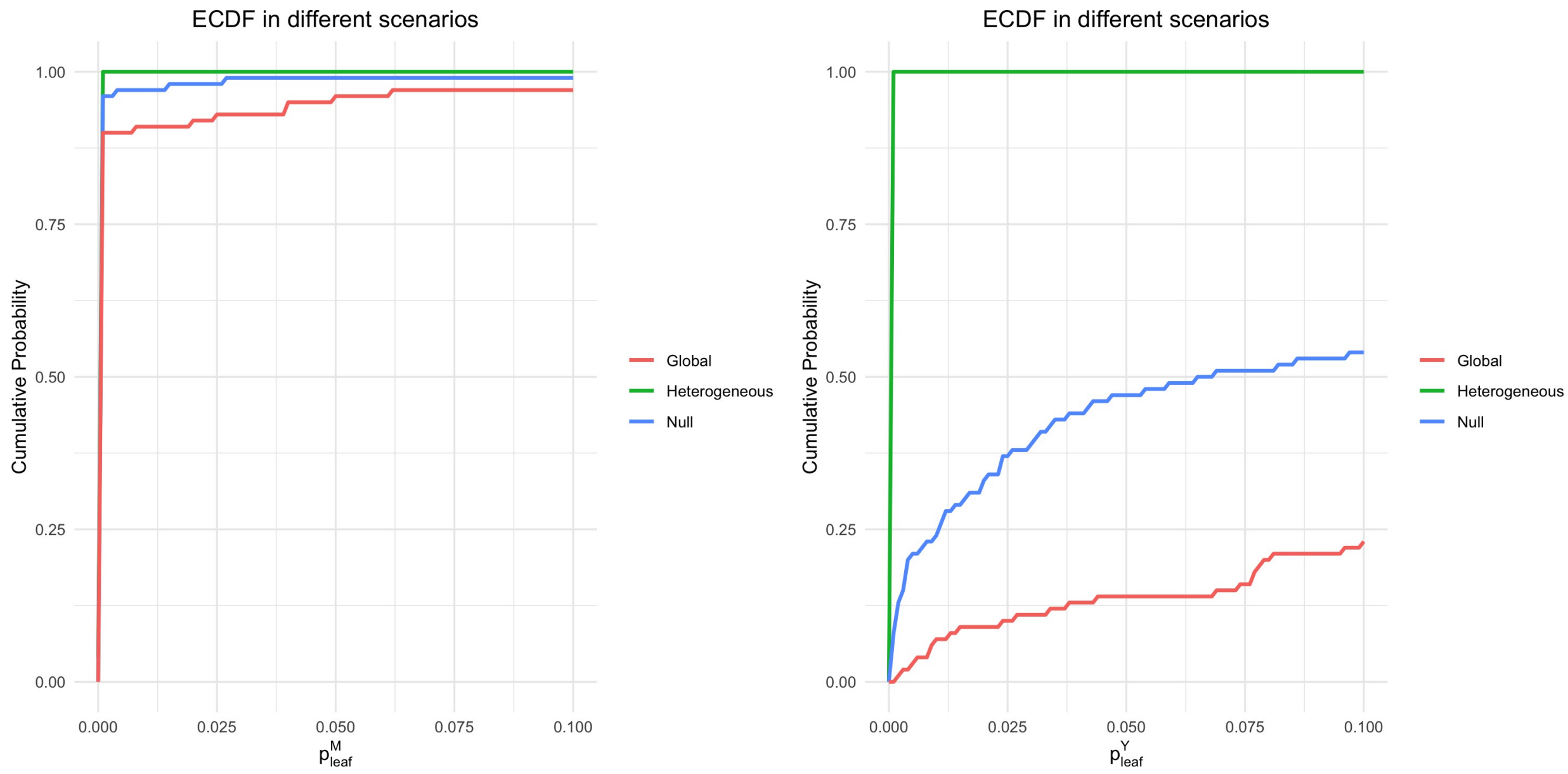

**Web Figure 3.** ECDF of linear assumption with RKLS method.

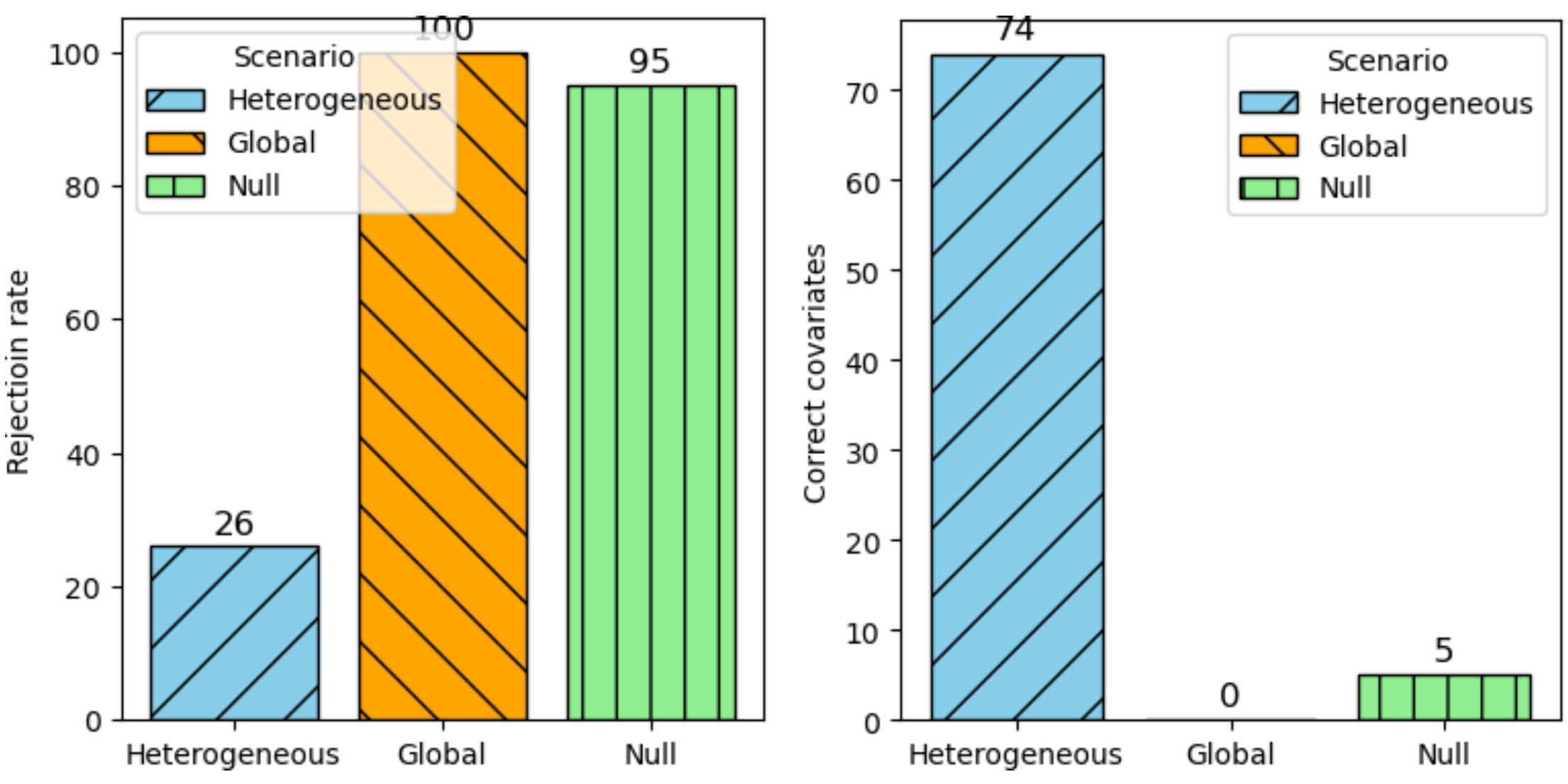

**Web Figure 4.** Left: the number of runs (out of 100) with no output under the linear assumption with RKLS method. Right: the frequency with which the final output profile selected $X^{(1)}$ and $X^{(2)}$ across the 100 runs with RKLS method.

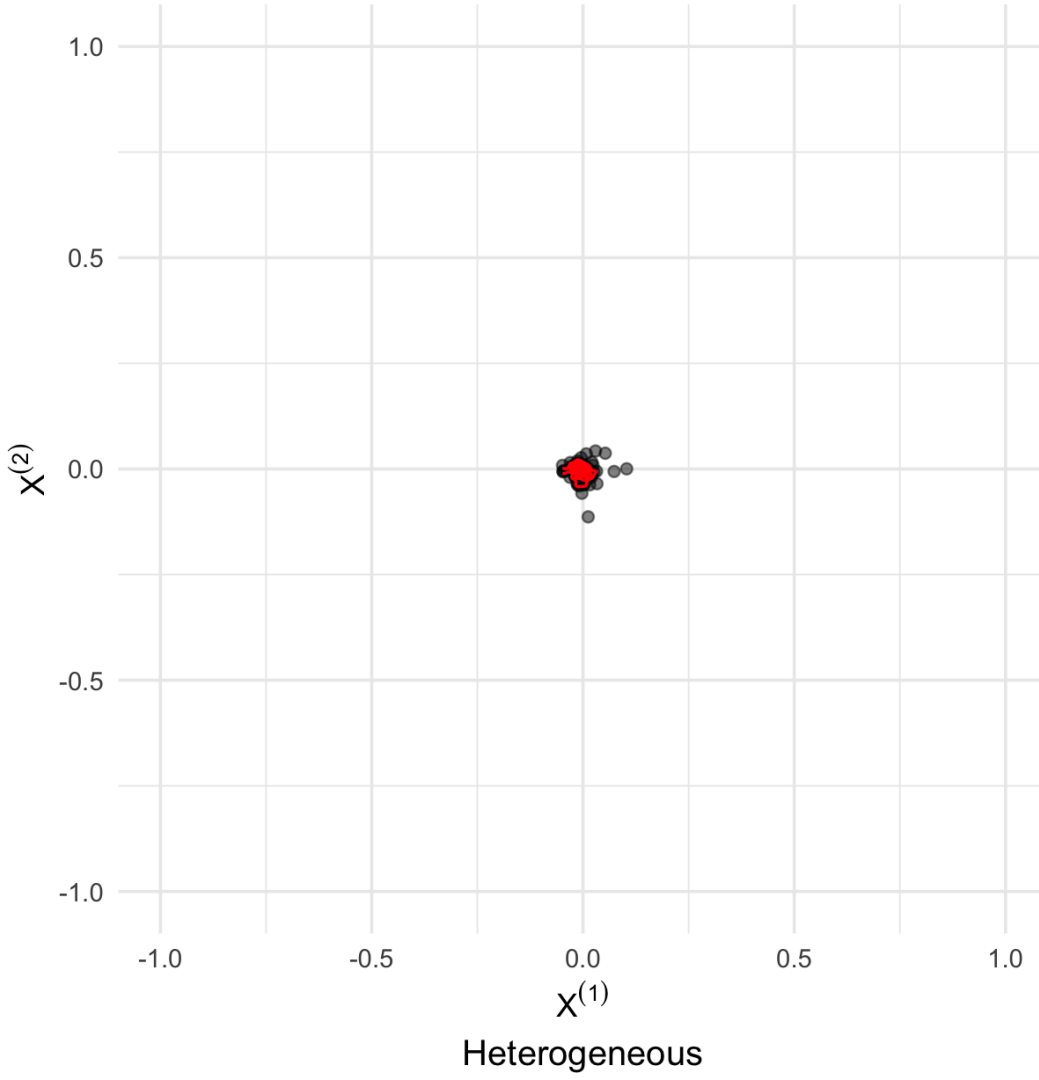

**Web Figure 5.** Distribution of thresholds of linear assumption in Heterogeneous scenario with random forest method. Black points denote replicate-specific estimates, and the red curves represent the corresponding density estimates.

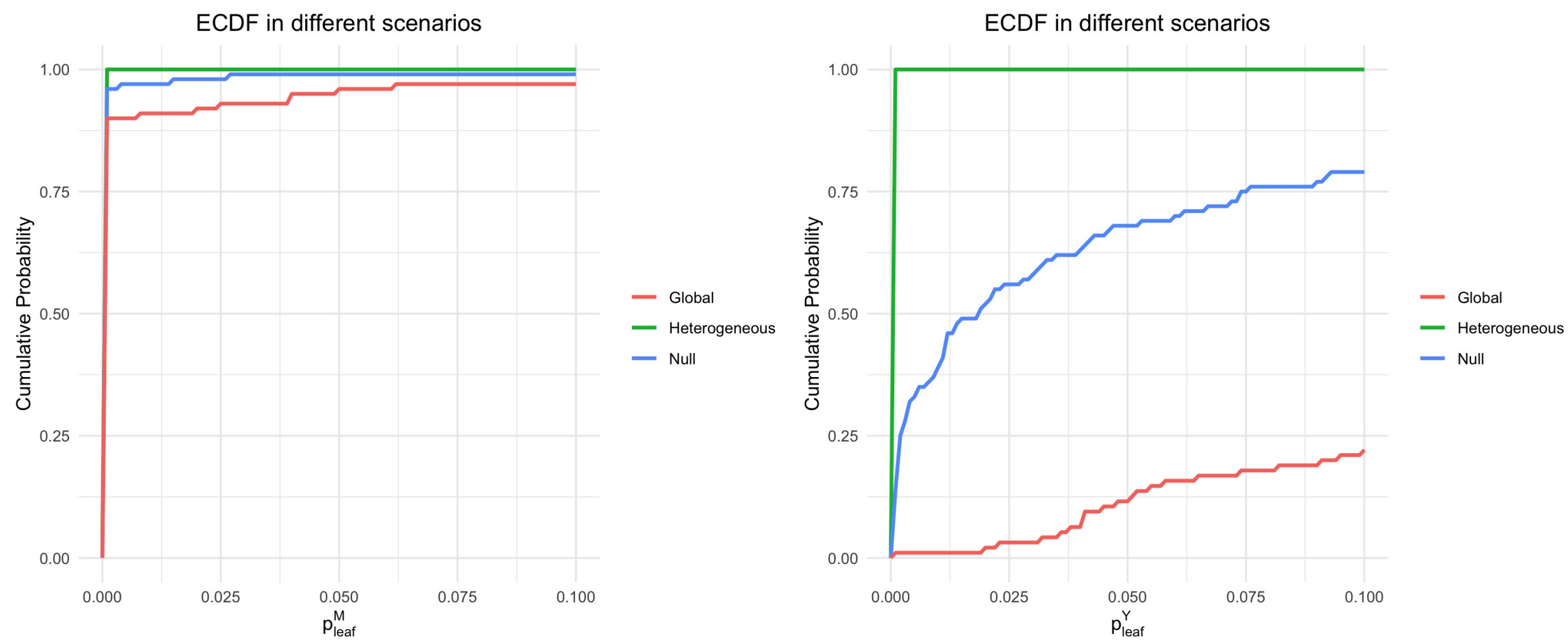

**Web Figure 6.** ECDF of linear assumption with random forest method.

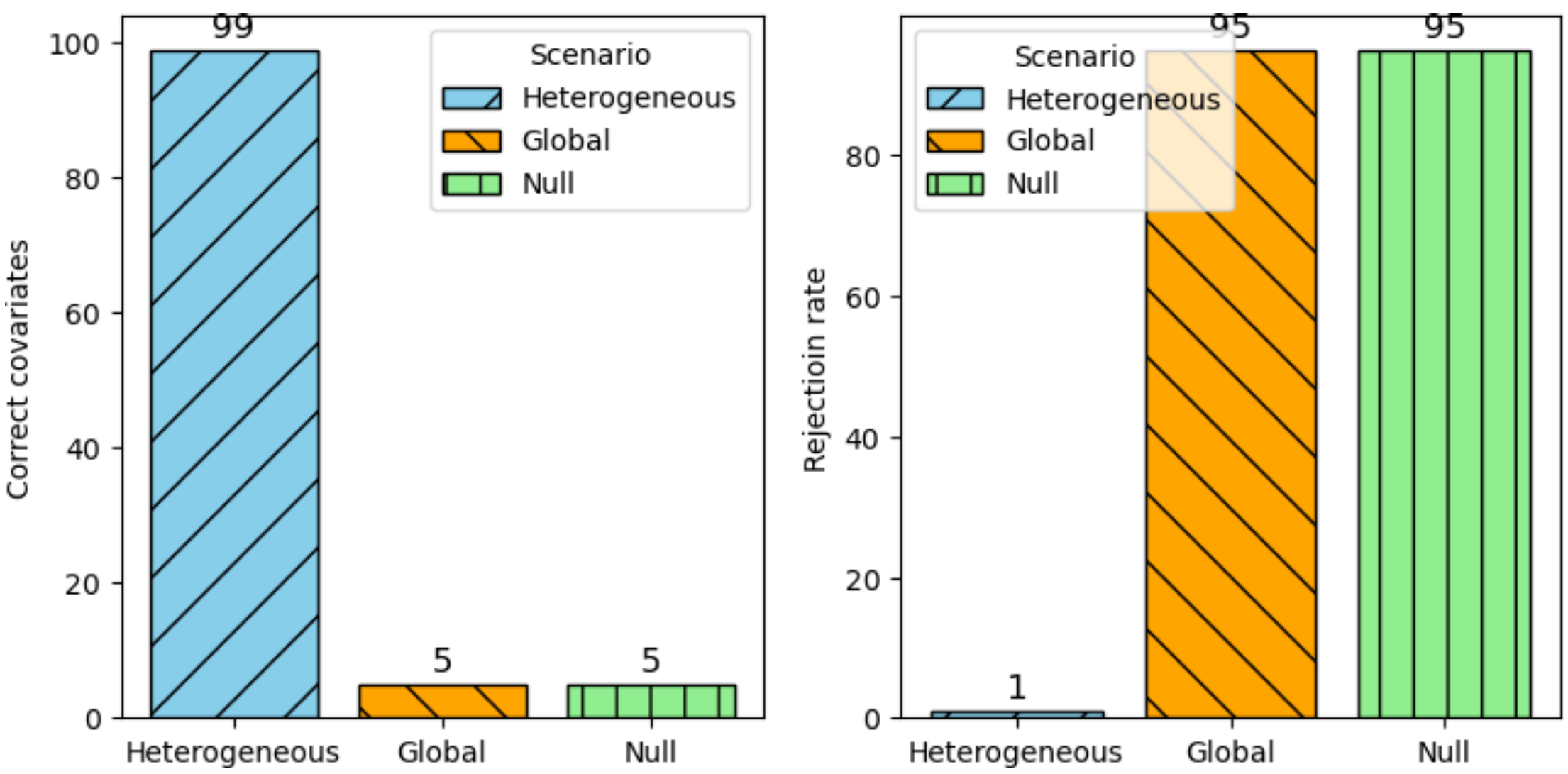

**Web Figure 7.** Left: the number of runs (out of 100) with no output under the linear assumption with random forest method. Right: the frequency with which the final output profile selected $X^{(1)}$ and $X^{(2)}$ across the 100 runs with random forest method.

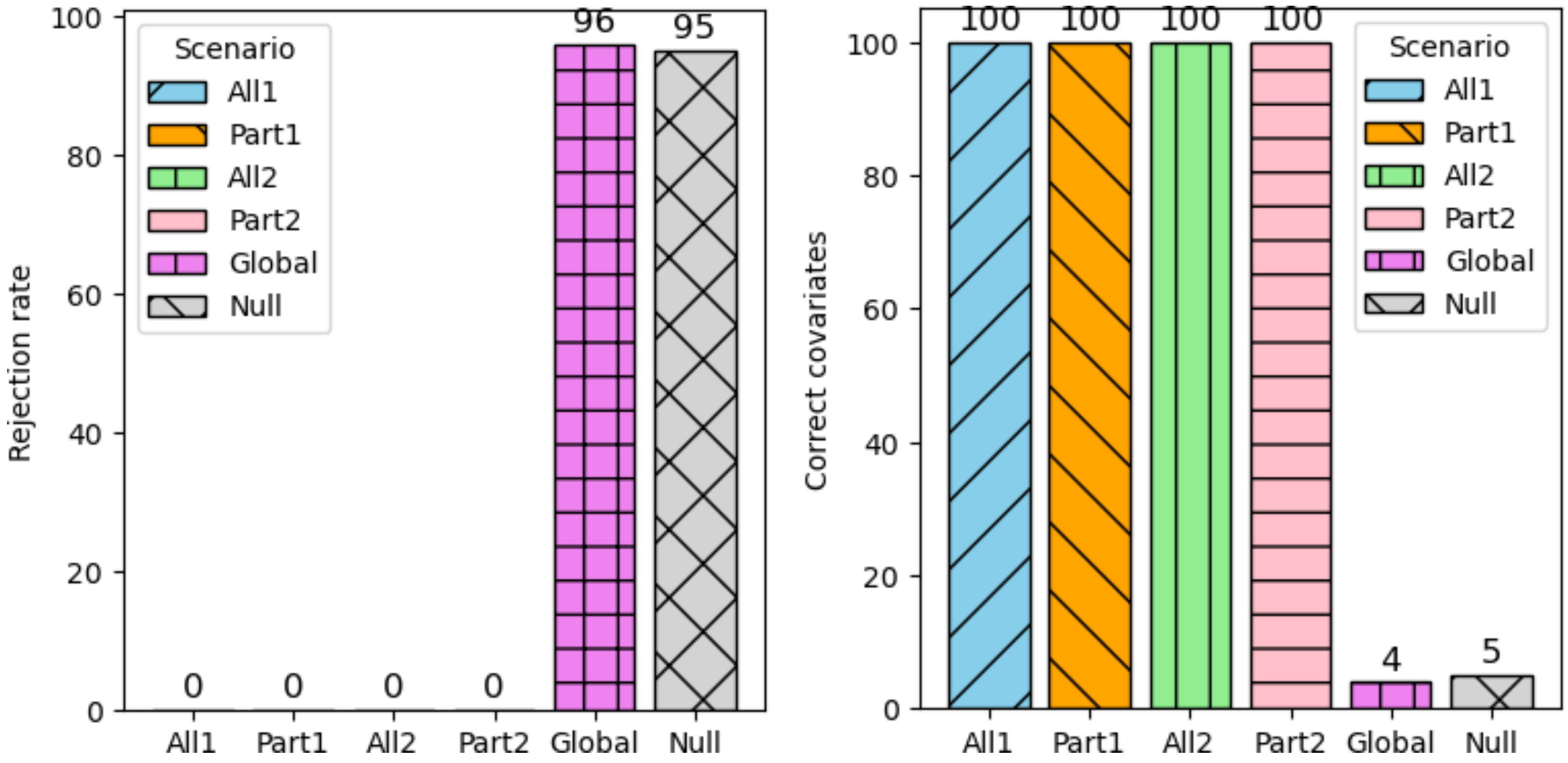

**Web Figure 8.** Left: the number of runs (out of 100) with no output under the complex assumption. Right: the frequency with which the final output profile selected $X^{(1)}$ and $X^{(2)}$ across the 100 runs. Scenarios All1 and All2 correspond to complete mediation of the treatment effect through the mediator, whereas Part1 and Part2 correspond to partial mediation. The treatment effect is constant in All1 and Part1, and covariate-dependent in All2 and Part2.

# 8 Web Appendix H: Additional results for real data analysis

For the HIV dataset, we first conducted the analysis under a linear model assumption. The mediator function was estimated using a random forest model, with the number of trees set to 1000, while all other parameters were kept at the default settings of the randomForestSRC package. The analysis was repeated six times with different random seeds, and the final result was selected according to a custom metric. The resulting subtype profile was shown in Web Figure 9, and the distribution of individual was shown in Web Figure 10.

The results indicate heterogeneity with respect to baseline CD4 levels. Individuals with baseline CD4 counts below 271 were classified as Subgroup 1, whereas those with baseline CD4 counts greater than or equal to 271 were classified as Subgroup 2. The average covariate-conditional indirect treatment effects for the two subtypes were $-0.53$ and $-0.33$, respectively.

*Remark* 4. The results from the linear assumption motivated us to further investigate the data under more complex modeling assumptions. The analysis of linear assumption suggested heterogeneity with respect to baseline CD4 levels; however, CD4 is a special variable in this study. At Week 0, CD4 serves as a baseline covariate, whereas at Week 20 it acts as a mediator that reflects the treatment effect transmitted through CD4. Consequently, the effect of CD4 on survival may vary across different CD4 levels, indicating potential nonlinear and heterogeneous indirect treatment effects.

## 8.1 Counterfactual survival plots

To construct the counterfactual survival plots, we proceeded as follows. Using the observed data $(T_i, \delta_i, M_i, X_i, W_i)$, we first fitted a Cox proportional hazards model with XGBoost to estimate the conditional hazard function

$$\lambda(t \mid W_i, M_i, X_i) = \lambda_0(t)\exp\{\eta(W_i, M_i, X_i)\},$$

where $\lambda_0(t)$ denotes the baseline hazard function and $\eta(\cdot)$ is a nonlinear function estimated by XGBoost.

Next, we fitted a mediator model describing the relationship between the mediator, treatment, and baseline covariates,

$$M_i = g(W_i, X_i) + \varepsilon_i.$$

Using the fitted mediator model, we generated the counterfactual mediator values by setting the treatment to $W = 1$ and $W = 0$, yielding

$$M_i(1), \qquad M_i(0).$$

These predicted mediator values were then substituted into the fitted survival model to estimate the counterfactual survival functions

$$S_i^{(1,M_i(1))}(t), \qquad S_i^{(1,M_i(0))}(t), \qquad S_i^{(0,M_i(0))}(t).$$

Finally, based on the three subtypes identified by our proposed method, we estimated and plotted the corresponding counterfactual survival curves for each subtype.

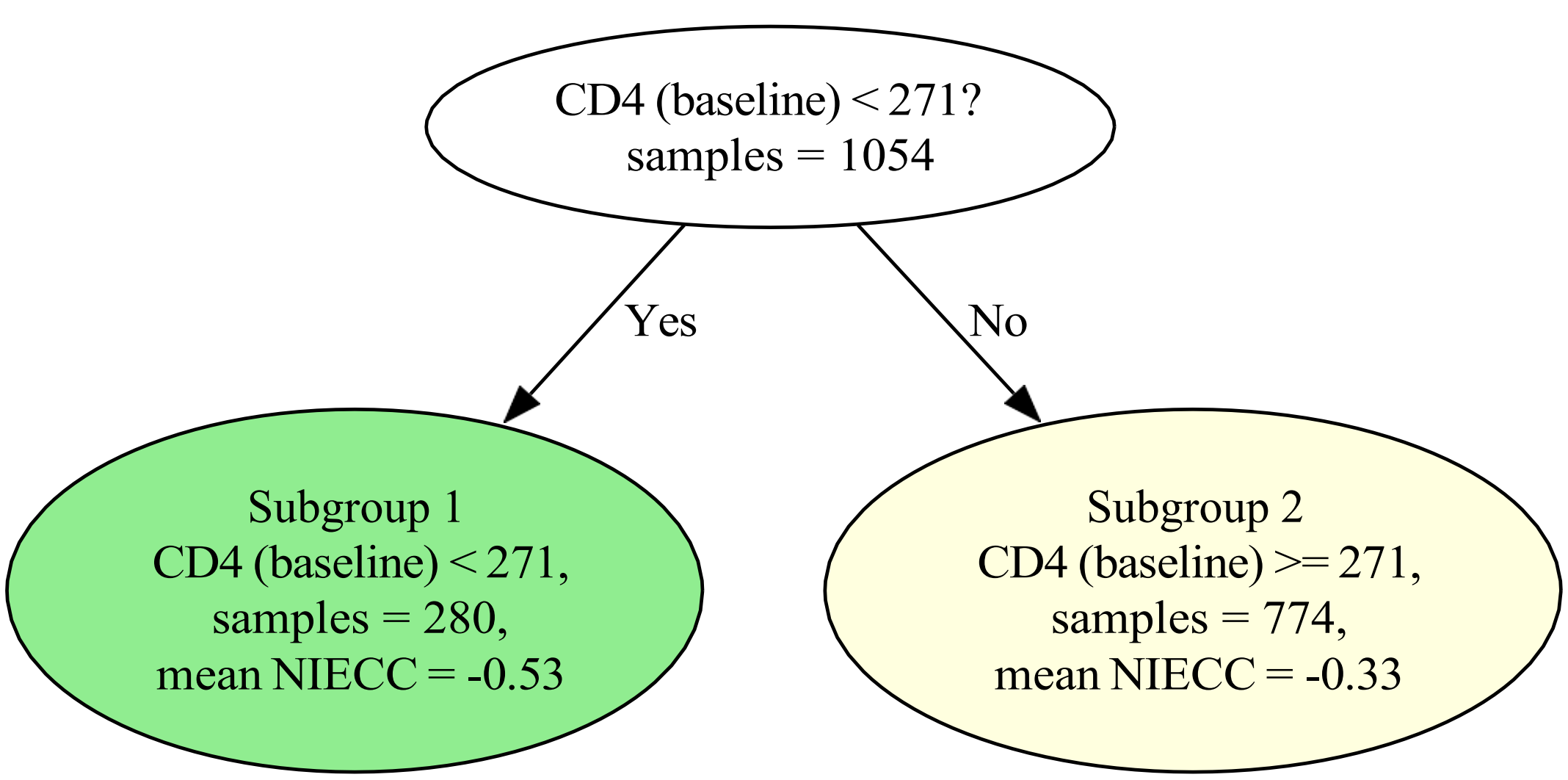

**Web Figure 9.** Profile of heterogeneous surrogate biomarker with linear assumption. mean NIECC: mean natural indirect treatment effect conditional on covariates in this group.

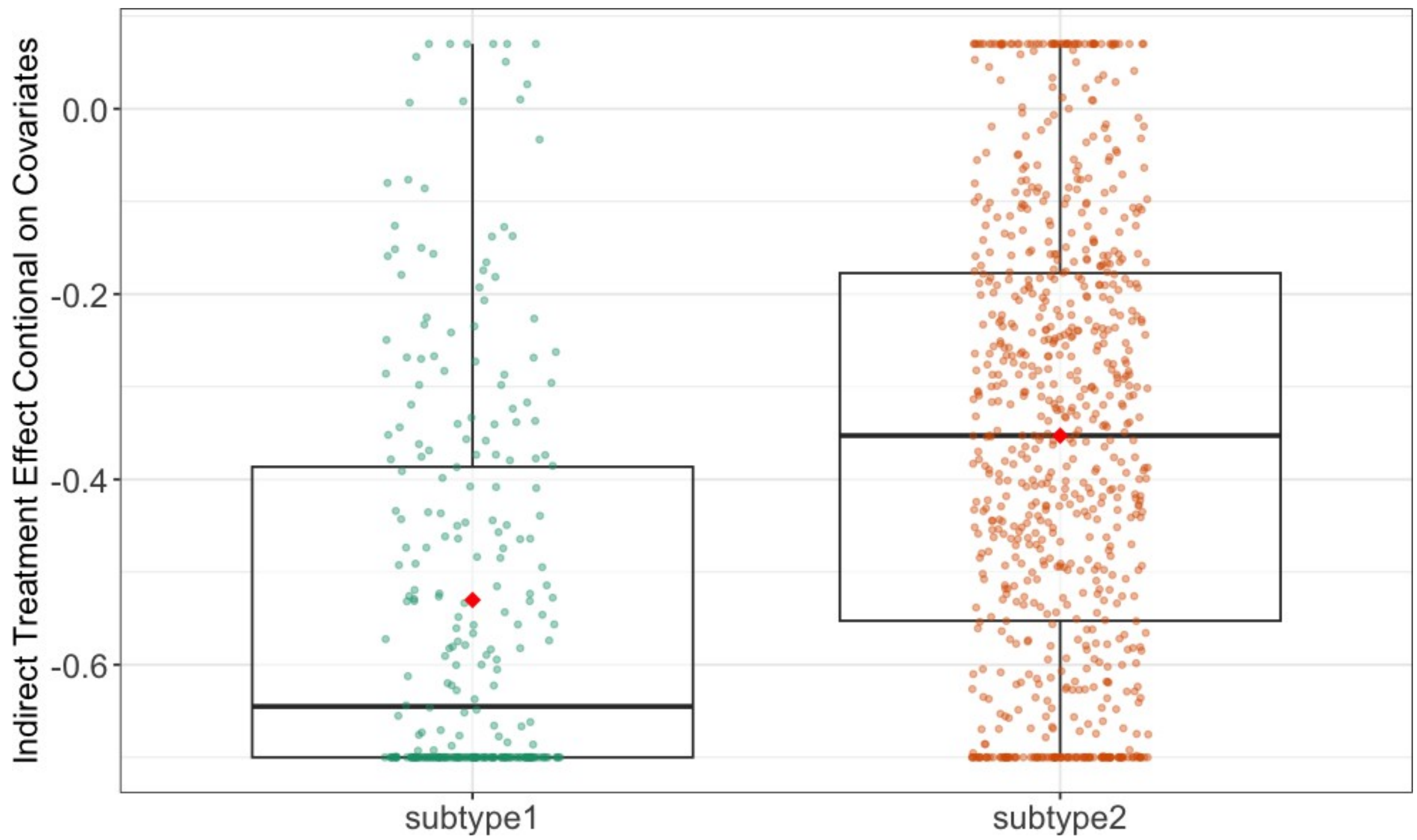

**Web Figure 10.** Distribution of NIECC across two subtypes with linear assumption. Each point represents an individual-level indirect treatment effect conditional on covariates. Boxes indicate the interquartile range with medians shown as horizontal lines, and red diamonds denote subgroup means. The mean indirect effects are $-0.53$ for Subtype 1, $-0.33$ for Subtype 2. The indirect treatment effect estimates are truncated by displaying only individuals whose estimated NIECC lie between the 10st and 90th percentiles.